\documentclass[fleqn,usenatbib]{mnras}

\usepackage{newtxtext,newtxmath}
\usepackage[T1]{fontenc}

\usepackage{tabu}
\usepackage{threeparttable, threeparttablex}
\usepackage{lscape}
\usepackage{pdflscape}
\usepackage{verbatim}
\usepackage{booktabs}
\usepackage{subfig}
\usepackage{stfloats}
\usepackage{gensymb}
\usepackage{bm}

\DeclareRobustCommand{\VAN}[3]{#2}
\let\VANthebibliography\thebibliography
\def\thebibliography{\DeclareRobustCommand{\VAN}[3]{##3}\VANthebibliography}

\usepackage{graphicx}
\usepackage{amsmath}	
\newcommand{\mjup}{\ensuremath{M_{\mathrm{Jup}}}} 
\newcommand{\nodata}{\centering\arraybackslash\cdot\cdot\cdot}
\newcommand{\rPS}{$r_{\rm P1}$}
\newcommand{\iPS}{$i_{\rm P1}$}
\newcommand{\zPS}{$z_{\rm P1}$}
\newcommand{\yPS}{$y_{\rm P1}$}
\newcommand{\Wone}{$W1$}
\newcommand{\Wtwo}{$W2$}
\newcommand{\Wthr}{$W3$}
\newcommand{\YMKO}{$Y_{\rm MKO}$}
\newcommand{\JMKO}{$J_{\rm MKO}$}
\newcommand{\HMKO}{$H_{\rm MKO}$}
\newcommand{\KMKO}{$K_{\rm MKO}$}
\newcommand{\JtwoM}{$J_{\rm 2M}$}
\newcommand{\HtwoM}{$H_{\rm 2M}$}
\newcommand{\KtwoM}{$K_{\rm 2M}$}
\newcommand{\Teff}{$T_{\rm eff}$}
\newcommand{\SpT}{\textit{SpT}}
\newcommand{\logL}{$\log L$}
\newcommand{\TC}{\textit{The Cannon}}
\newcommand{\HtwoO}{H$_2$O}
\newcommand{\CHfour}{CH$_4$}

\title[Determining Physical Properties of Brown Dwarfs]{Using Old and New Approaches: Determining Physical Properties of Brown Dwarfs with Empirical Relations and Machine Learning Models}

\author[Feeser, Best]{
S. Jean Feeser,$^{1}$\thanks{E-mail: sjfeeser@utexas.edu}
William M. J. Best,$^{1}$\thanks{E-mail: wbest@utexas.edu}\\
$^{1}$University of Texas at Austin, Department of Astronomy, 2515 Speedway C1400, Austin, TX 78712, USA\\
}

\date{Accepted XXX. Received YYY; in original form ZZZ}

\pubyear{2022}

\begin{document}
\label{firstpage}
\pagerange{\pageref{firstpage}--\pageref{lastpage}}
\maketitle

\begin{abstract}
We investigate applications of machine learning models to directly infer physical properties of brown dwarfs from their photometry and spectra using {\TC}. We demonstrate that absolute magnitudes, spectral types, and spectral indices can be determined from low-resolution SpeX prism spectra of L and T dwarfs without trigonometric parallax measurements and with precisions competitive with commonly used methods. For T~dwarfs with sufficiently precise spectra and photometry, bolometric luminosities and effective temperatures can be determined at precisions comparable to methods that use polynomial relations as a function of absolute magnitudes. We also provide new and updated polynomial relations for absolute magnitudes as a function of spectral types L0--T8 in 14 bands spanning Pan-STARRS {\rPS} to AllWISE {\Wthr}, using a volume-limited sample of 256 brown dwarfs defined entirely by parallaxes. These include the first relations for brown dwarfs using Pan-STARRS1 photometry and the first for several infrared bands using a volume-limited sample. We find that our novel method with {\TC} can infer absolute magnitudes with equal or smaller uncertainties than the polynomial relations that depend on trigonometric parallax measurements.
\end{abstract}


\section{Introduction}

Brown dwarfs are self-gravitating objects more massive than planets {\citep[$\gtrsim$13~\mjup;][]{Spiegel2011}} whose central temperatures and densities are not sufficient for hydrogen fusion to occur {\citep[$\lesssim$70~\mjup;][]{Dupuy2017}}. These faint, substellar objects cool continuously as they age through L ($\approx$2000~K), T ($\approx$1000~K), and Y ($\approx$300~K) spectral types. Spectroscopic analysis provides a useful tool for classification of brown dwarfs and is crucial for discerning their physical properties such as luminosity, temperature, and surface gravity. However, determining these parameters is complicated by their evolution through cooler temperatures and the challenges associated with characterizing precise atmospheric properties from spectra, limiting the information we can extract from modelling and fitting techniques. 

Absolute magnitudes of brown dwarfs are also essential data for constraining and modelling evolution and physical properties, and for flux-calibrating spectra. However, absolute magnitudes require distance measurements that are observationally expensive for the many brown dwarfs that are too faint and red for \emph{Gaia} \citep{Best2020a}. A more efficient method to determine absolute magnitudes is to use empirical relationships with spectral types calibrated by objects with existing parallax measurements  \citep[e.g.,][]{Dupuy2012,Faherty2016a,Kirkpatrick2021}. (In this paper, ``parallax'' refers to trigonometric parallax.) In Section~\ref{sec:poly} we present updated polynomial relations between spectral type and absolute magnitude based on a volume-limited sample in 14 bands. These include the first determined for Panoramic Survey Telescope and Rapid Response System (Pan-STARRS1) 3$\pi$ Survey \citep[hereinafter PS1;][]{Chambers2016} photometry and the first using a volume-limited sample for the Two Micron All Sky Survey \citep[2MASS;][]{Skrutskie2006} photometric system in $J$ and $K$ bands, $Y$ and $K$ bands for the Mauna Kea Observatories (MKO) photometric system \citep{Simons:2002hh,Tokunaga:2002ex}, and the Wide-Field Infrared Survey Explorer \citep[AllWISE;][]{Wright2010} photometric system. Spectral types, however, are heterogeneously determined and do not capture subtle differences in spectral features between objects, limiting the accuracy of these polynomial relations.

Machine learning has been used in astronomical research as a tool to process, analyse, and extract new information from large amounts of data \citep{Ivezic2014}. For example, machine learning models such as {\TC} \citep{Ness2015,Ho2017a} have been utilized to infer stellar parameters and abundances from spectroscopic data. These models employ ``stellar labels'' (i.e. measured quantities) from reference stars to infer unknown labels for a new sample of objects. Previous applications of {\TC} have used reliable measurements of physical quantities such as effective temperature, surface gravity, and chemical abundances \citep{Casey2017, Behmard2019}, but any ``known'' labels could be used to acquire new information from the spectrum of an object.  Proven to work well at low signal-to-noise, {\TC} could become an effective means of obtaining new information from the spectra of brown dwarfs. 
(We note also that \citealt{Aganze:2021} recently used alternative machine learning methods to identify late-M, L, and T dwarfs in large, low-resolution spectroscopic surveys. Their goal was to classify approximate spectral types, rather than physical parameters, but they demonstrated clear success in distinguishing low-S/N ultracool dwarf spectra from contaminants, e.g., galaxies, using both Random Forests and Neural Networks.)
In Section~\ref{sec:TC} we present the results of our investigation into this novel approach to analysing brown dwarf spectra using data-driven models and discuss the efficacy of {\TC} for future applications.

\section{Spectral Type Polynomials}
\label{sec:poly}

\subsection{The Volume-Limited Sample}
\label{sec:sample}

Volume-limited samples are ideal for stellar population studies because they minimize selection biases. Starting with a sample of over 1200 L, T, and Y dwarfs with parallax measurements and PS1, MKO, 2MASS, and/or AllWISE photometry compiled from the literature, we extracted objects with declinations $-30\degree \leq \delta \leq 60\degree$ and parallax-determined distances less than 25 pc, replicating the volume-limited sample of \citet{Best2021} but also including brown dwarfs with spectral types T8.5--Y1.

Brown dwarfs with ages $\lesssim$200~Myr are still undergoing significant contraction and will exhibit inflated radii when compared to older brown dwarfs \citep{Burrows1997}. These young objects will therefore have relatively low surface gravity and photospheric pressure, causing them to differ spectroscopically and photometrically from field dwarfs \citep[e.g.,][]{LiuDupuy2016}. Older objects with lower metallicities will also differ from field objects, appearing bluer \citep{Kirkpatrick2008}. Therefore, we removed any object classified as young or a subdwarf by \citet{Best2021}. 
Unresolved binaries that blend the spectra of two object can exhibit unusual spectroscopic features \citep[e.g.,][]{Burgasser2010}, so we also excluded known binaries, as well as objects that have not been confirmed as single using high-angular resolution imaging. Objects with  spectral types indicating peculiarities or unusually blue or red colours were additionally excluded.

Tables~\ref{tab:sample1}, \ref{tab:sample2}, and \ref{tab:sample3} show the volume-limited sample and photometry we used. The numerical spectral type (SpT) represents the adopted value used for the analysis, with $\text{L}0=10$ and $\text{Y}1=31$. We used optical spectral types for L dwarfs when available and near-infrared (NIR) otherwise; NIR spectral types were adopted for T and Y dwarfs. The final volume-limited sample contains 256 field objects; not all objects had photometry in all bands we analysed. For each band, we selected objects with parallax uncertainties $\leq$20~per~cent and absolute magnitude uncertainties $<0.2$~mag. Some bluer bands, primarily {\rPS} and {\iPS}, have limited samples at later spectral types which restricted our analysis in these ranges.

\begin{table*}
    \centering
    \caption{Volume-Limited Sample of Ultracool Dwarfs Used For Spectral Type to Absolute Magnitude Polynomials}
    \label{tab:sample1}
    \begin{tabular}{lcccccc}
         \toprule
         \toprule
         Object & Parallax & RA & Dec & Spectral Type & Numerical SpT & References\\
          & (mas) & (deg) & (deg) & (Opt./NIR) & (Adopted) & (Disc.; Plx.; SpT)\\
         \cmidrule(lr){1-7}
         SDSS J000013.54+255418.6 & 70.8 $\pm$ 1.9 & 0.0564 & 25.905389 & T5/T4.5 & 24.5 & 68; 41; 92,15 \\
         WISE J000517.48+373720.5 & 127.0 $\pm$ 2.4 & 1.322947 & 37.622348 & $\nodata$/T9 & 29.0 & 83; 66; 83 \\
         2MASS J00132229$-$1143006 & 40.3 $\pm$ 3.1 & 3.343632 & $-$11.716855 & $\nodata$/T4 & 24.0 & 56; 6; 6 \\
         2MASSW J0015447+351603 & 58.6 $\pm$ 0.4 & 3.936774 & 35.266588 & L2/L1.0 & 12.0 & 60; 46; 60,3 \\
         PSO J004.6359+56.8370 & 46.5 $\pm$ 3.9 & 4.636015 & 56.837043 & $\nodata$/T4.5 & 24.5 & 6; 6; 6 \\
         PSO J007.9194+33.5961 & 45.4 $\pm$ 3.8 & 7.919394 & 33.596035 & $\nodata$/L9 & 19.0 & 5; 6; 5 \\
         2MASS J00320509+0219017 & 41.0 $\pm$ 0.4 & 8.022519 & 2.31609 & L1.5/M9 & 11.5 & 97; 46; 97,112 \\
         ULAS J003402.77$-$005206.7 & 68.7 $\pm$ 1.4 & 8.511645 & $-$0.86896 & $\nodata$/T9 & 29.0 & 111; 41; 20 \\
         2MASS J00345157+0523050 & 120.1 $\pm$ 3.0 & 8.716832 & 5.385247 & $\nodata$/T6.5 & 26.5 & 14; 66; 15 \\
         WISE J003829.05+275852.1 & 89.7 $\pm$ 2.5 & 9.621042 & 27.981139 & $\nodata$/T9 & 29.0 & 83; 66; 83 \\
         HD 3651B & 89.8 $\pm$ 0.1 & 9.827351 & 21.253335 & $\nodata$/T7.5 & 27.5 & 90; 46; 81 \\
         WISE J004024.88+090054.8 & 69.8 $\pm$ 1.5 & 10.103758 & 9.01515 & $\nodata$/T7 & 27.0 & 83; 66; 83 \\
         WISE J004542.56+361139.1 & 53.4 $\pm$ 5.2 & 11.427445 & 36.194244 & $\nodata$/T5 & 25.0 & 83; 6; 83 \\
         WISEPC J004928.48+044100.1 & 62.6 $\pm$ 2.9 & 12.368717 & 4.683329 & $\nodata$/L9 & 19.0 & 63; 6; 63 \\
         WISE J004945.61+215120.0 & 139.9 $\pm$ 2.5 & 12.440559 & 21.85549 & $\nodata$/T8.5 & 28.5 & 83; 66; 83 \\
         CFBDS J005910.90$-$011401.3 & 103.2 $\pm$ 2.1 & 14.794744 & $-$1.233656 & $\nodata$/T8.5 & 28.5 & 40; 41; 30 \\
         \bottomrule
    \end{tabular}
    \begin{tablenotes}
            \item \textbf{Notes.} This table contains 256 single L0--Y1 dwarfs with declinations between $-30\degree$ and $+60\degree$ and parallaxes greater than 40~mas (i.e., distances less than 25~pc), which we used to determine our spectral type vs. absolute photometry polynomials. The numerical spectral type (SpT) represents the value adopted for the analysis, with $\text{L}0=10$ and $\text{Y}1=31$. The table is available in its entirety in a machine-readable form in the online journal. A portion is shown here.
            \item \textbf{References.} {(1) \citet{Albert:2011bc}, (2) \citet{Allers2013}, (3) \citet{BardalezGagliuffi2014}, (4) \citet{Best:2013bp}, (5) \citet{Best:2015em}, (6) \citet{Best2020a}, (7) \citet{Bihain:2013gw}, (8) \citet{Burgasser:1999fp}, (9) \citet{Burgasser:2000bm}, (10) \citet{Burgasser:2002fy}, (11) \citet{Burgasser:2003dh}, (12) \citet{Burgasser:2003ij}, (13) \citet{Burgasser:2003jf}, (14) \citet{Burgasser:2004hg}, (15) \citet{Burgasser2006}, (16) \citet{Burgasser:2008cj}, (17) \citet{Burgasser:2008ei}, (18) \citet{Burgasser:2008ke}, (19) \citet{Burgasser2010}, (20) \citet{Burningham:2008fc}, (21) \citet{Burningham:2009ft}, (22) \citet{Burningham:2010dh}, (23) \citet{Burningham:2011kh}, (24) \citet{Burningham:2013gt}, (25) \citet{Castro:2012dj}, (26) \citet{Castro:2013bb}, (27) \citet{Chiu:2006jd}, (28) \citet{Cruz:2003fi}, (29) \citet{Cruz:2007kb}, (30) \citet{Cushing:2011dk}, (31) \citet{Cushing:2014be}, (32) \citet{Dahn:2017gu}, (33) \citet{Deacon:2011gz}, (34) \citet{Deacon:2012eg}, (35) \citet{Deacon:2012gf}, (36) \citet{Deacon:2014ey}, (37) \citet{Deacon:2017kd}, (38) \citet{Delfosse:1997uj}, (39) \citet{Delfosse:1999bx}, (40) \citet{Delorme:2008jd}, (41) \citet{Dupuy2012}, (42) T. Dupuy (private communication), (43) \citet{Faherty:2012cy}, (44) \citet{Fan:2000iu}, (45) \citet{Gagne:2015dc}, (46) \citet{GaiaCollaboration:2018io}, (47) \citet{Geballe:2002kw}, (48) \citet{Gizis:2000kz}, (49) \citet{Gizis:2001jp}, (50) \citet{Gizis:2002je}, (51) \citet{Gizis:2011fq}, (52) \citet{Gizis:2011jv}, (53) \citet{Gizis:2013ik}, (54) \citet{Gizis:2015fa}, (55) \citet{Hawley:2002jc}, (56) \citet{Kellogg:2017kh}, (57) \citet{Kendall:2004kb}, (58) \citet{Kendall:2007fd}, (59) \citet{Kirkpatrick:1999ev}, (60) \citet{Kirkpatrick:2000gi}, (61) \citet{Kirkpatrick2008}, (62) \citet{Kirkpatrick:2010dc}, (63) \citet{Kirkpatrick:2011ey}, (64) \citet{Kirkpatrick:2012ha}, (65) \citet{Kirkpatrick:2014kv}, (66) \citet{Kirkpatrick:2019kt}, (67) \citet{Kirkpatrick2021}, (68) \citet{Knapp2004a}, (69) \citet{Leggett:2000ja}, (70) \citet{Leggett:2012gg}, (71) \citet{Liebert:2003bx}, (72) \citet{Liu:2002fx}, (73) \citet{Liu:2011hc}, (74) \citet{Liu:2013gy}, (75) \citet{Lodieu:2005kd}, (76) \citet{Lodieu:2007fr}, (77) \citet{Lodieu:2012go}, (78) \citet{Looper:2007ee}, (79) \citet{Looper:2008hs}, (80) \citet{Lucas:2010iq}, (81) \citet{Luhman:2007fu}, (82) \citet{Luhman:2012ir}, (83) \citet{Mace:2013jh}, (84) \citet{Manjavacas:2013cg}, (85) \citet{Marocco:2010cj}, (86) \citet{Marocco:2013kv}, (87) \citet{Marocco:2015iz}, (88) \citet{Martin:2010cx}, (89) \citet{Martin:2018hc}, (90) \citet{Mugrauer:2006iy}, (91) \citet{PhanBao:2008kz}, (92) \citet{Pineda:2016ku}, (93) \citet{Pinfield:2008jx}, (94) \citet{Pinfield:2012hm}, (95) \citet{Radigan:2008jd}, (96) \citet{Reid:2000iw}, (97) \citet{Reid:2008fz}, (98) \citet{Sahlmann:2014hu}, (99) \citet{Schneider:2014jd}, (100) \citet{Schneider:2015bx}, (101) \citet{Scholz:2002by}, (102) \citet{Scholz:2010cy}, (103) \citet{Scholz:2011gs}, (104) \citet{Smart:2018en}, (105) \citet{Strauss:1999iw}, (106) \citet{Thompson:2013kv}, (107) \citet{Tinney:2003eg}, (108) \citet{Tinney:2005hz}, (109) \citet{Tsvetanov:2000cg}, (110) \citet{Vrba:2004ee}, (111) \citet{Warren:2007kw}, (112) \citet{Wilson:2003tk}, (113) \citet{Wright:2013bo}, (114) \citet{vanLeeuwen:2007dc}.}
    \end{tablenotes}
\end{table*}

\begin{table*}
    \centering
    \caption{Absolute MKO and 2MASS Photometry for the Volume-Limited Sample}
    \label{tab:sample2}
    \begin{tabular}{lcccccccc}
         \toprule
         \toprule
         \multicolumn{1}{c}{} & \multicolumn{4}{c}{MKO} & \multicolumn{3}{c}{2MASS}\\
         \cmidrule(lr){2-5}\cmidrule(lr){6-8}
         Object & $Y$ & $J$ & $H$ & $K$ & $J$ & $H$ & $K$ & References\\
          & (mag) & (mag) & (mag) & (mag) & (mag) & (mag) & (mag)\\
         \cmidrule(lr){1-9}
         SDSS J000013.54+255418.6 & 15.05 $\pm$ 0.08 & 13.98 $\pm$ 0.07 & 13.99 $\pm$ 0.07 & 14.07 $\pm$ 0.07 & 14.31 $\pm$ 0.07 & 13.98 $\pm$ 0.13 & 14.09 $\pm$ 0.13 & 14,21,27 \\
         WISE J000517.48+373720.5 & 19.00 $\pm$ 0.05 & 18.11 $\pm$ 0.05 & 18.5 $\pm$ 0.05 & 18.51 $\pm$ 0.05 & $\nodata$ & $\nodata$ & $\nodata$ & 34 \\
         2MASS J00132229$-$1143006 & $\nodata$ & 14.08 $\pm$ 0.17 & $\nodata$ & $\nodata$ & 14.38 $\pm$ 0.19 & 13.71 $\pm$ 0.28 & 13.72 $\pm$ 0.28 & 14,5 \\
         2MASSW J0015447+351603 & 13.79 $\pm$ 0.06 & 12.58 $\pm$ 0.02 & 11.80 $\pm$ 0.03 & 11.09 $\pm$ 0.03 & 12.72 $\pm$ 0.03 & 11.73 $\pm$ 0.02 & 11.10 $\pm$ 0.02 & 14,6,5 \\
         PSO J004.6359+56.8370 & $\nodata$ & 14.56 $\pm$ 0.18 & $\nodata$ & $\nodata$ & 14.82 $\pm$ 0.24 & $\nodata$ & $\nodata$ & 14,5 \\
         PSO J007.9194+33.5961 & 15.77 $\pm$ 0.19 & 14.67 $\pm$ 0.18 & 13.74 $\pm$ 0.19 & 12.96 $\pm$ 0.19 & 14.74 $\pm$ 0.22 & 13.72 $\pm$ 0.2 & 12.78 $\pm$ 0.20 & 14,6,5 \\
         2MASS J00320509+0219017 & 13.51 $\pm$ 0.02 & 12.29 $\pm$ 0.02 & 11.51 $\pm$ 0.02 & 10.86 $\pm$ 0.02 & 12.39 $\pm$ 0.04 & 11.46 $\pm$ 0.04 & 10.87 $\pm$ 0.04 & 14,28 \\
         ULAS J003402.77$-$005206.7 & 18.08 $\pm$ 0.11 & 17.33 $\pm$ 0.05 & 17.67 $\pm$ 0.06 & 17.66 $\pm$ 0.07 & $\nodata$ & $\nodata$ & $\nodata$ & 50 \\
         2MASS J00345157+0523050 & 16.61 $\pm$ 0.05 & 15.54 $\pm$ 0.05 & 15.97 $\pm$ 0.06 & 16.47 $\pm$ 0.06 & 15.94 $\pm$ 0.07 & $\nodata$ & $\nodata$ & 14,28 \\
         WISE J003829.05+275852.1 & $\nodata$ & 18.37 $\pm$ 0.06 & 18.68 $\pm$ 0.07 & $\nodata$ & $\nodata$ & $\nodata$ & $\nodata$ & 40 \\
         HD 3651B & 16.89 $\pm$ 0.06 & 15.93 $\pm$ 0.03 & 16.45 $\pm$ 0.04 & 16.64 $\pm$ 0.05 & 16.37 $\pm$ 0.03 & 16.43 $\pm$ 0.03 & 16.5 $\pm$ 0.03 & 21,6,39 \\
         WISE J004024.88+090054.8 & 16.37 $\pm$ 0.05 & 15.35 $\pm$ 0.05 & 15.78 $\pm$ 0.05 & 15.77 $\pm$ 0.07 & 15.72 $\pm$ 0.13 & $\nodata$ & $\nodata$ & 14,28 \\
         WISE J004542.56+361139.1 & 15.45 $\pm$ 0.22 & 14.55 $\pm$ 0.21 & 14.77 $\pm$ 0.22 & 14.71 $\pm$ 0.22 & 14.79 $\pm$ 0.23 & 14.68 $\pm$ 0.29 & 4.44 $\pm$ 0.29 & 14,6,5 \\
         WISEPC J004928.48+044100.1 & 15.89 $\pm$ 0.10 & 14.75 $\pm$ 0.10 & 13.79 $\pm$ 0.10 & 13.11 $\pm$ 0.10 & 14.83 $\pm$ 0.12 & 13.65 $\pm$ 0.12 & 13.15 $\pm$ 0.12 & 14,28 \\
         WISE J004945.61+215120.0 & 18.02 $\pm$ 0.17 & 17.17 $\pm$ 0.16 & 17.45 $\pm$ 0.04 & 17.53 $\pm$ 0.17 & 17.45 $\pm$ 0.16 & $\nodata$ & $\nodata$ & 14,6,26 \\
         CFBDS J005910.90$-$011401.3 & 18.89 $\pm$ 0.05 & 18.13 $\pm$ 0.05 & 18.34 $\pm$ 0.07 & 18.78 $\pm$ 0.07 & 18.41 $\pm$ 0.05 & 18.27 $\pm$ 0.07 & 18.7 $\pm$ 0.07 & 21,20,32 \\
         \bottomrule
    \end{tabular}
    \begin{tablenotes}
            \item \textbf{Notes.} Absolute MKO and 2MASS photometry for all objects in the volume-limited sample (Table~\ref{tab:sample1}), calculated using the parallaxes in Table~\ref{tab:sample1}. This table is available in its entirety in a machine-readable form in the online journal. A portion is shown here.
            \item \textbf{References.} {(1) \citet{Albert:2011bc}, (2) \citet{Beichman:2014jr}, (3) \citet{Best:2013bp}, (4) \citet{Best:2015em}, (5) \citet{Best2020a}, (6) \citet{Best2021}, (7) \citet{Boccaletti:2003cl}, (8) \citet{Burningham:2008fc}, (9) \citet{Burningham:2009ft}, (10) \citet{Burningham:2010dh}, (11) \citet{Burningham:2013gt}, (12) \citet{Chiu:2006jd}, (13) \citet{Cushing:2014be}, (14) \citet{Cutri:2003vr}, (15) \citet{Deacon:2011gz}, (16) \citet{Deacon:2012eg}, (17) \citet{Deacon:2012gf}, (18) \citet{Deacon:2014ey}, (19) \citet{Deacon:2017kd}, (20) \citet{Delorme:2008jd}, (21) \citet{Dupuy2012}, (22) \citet{Dupuy:2013ks}, (23) \citet{Faherty:2012cy}, (24) \citet{Kirkpatrick:2011ey}, (25) \citet{Kirkpatrick:2012ha}, (26) \citet{Kirkpatrick:2019kt}, (27) \citet{Knapp2004a}, (28) \citet{Lawrence:2012wh}, (29) \citet{Leggett:2000ja}, (30) \citet{Leggett:2002cd}, (31) \citet{Leggett:2009jf}, (32) \citet{Leggett:2010cl}, (33) \citet{Leggett:2013dq}, (34) \citet{Leggett:2015dn}, (35) \citet{Leggett:2016fq}, (36) \citet{Lodieu:2012go}, (37) \citet{Lucas:2010iq}, (38) \citet{Lucas:2012wf}, (39) \citet{Luhman:2007fu}, (40) \citet{Mace:2013jh}, (41) \citet{Manjavacas:2013cg}, (42) \citet{McMahon:2013vw}, (43) \citet{PenaRamirez:2015id}, (44) \citet{Pinfield:2008jx}, (45) \citet{Pinfield:2012hm}, (46) \citet{Schneider:2015bx}, (47) \citet{Skrutskie2006}, (48) \citet{Strauss:1999iw}, (49) \citet{Thompson:2013kv}, (50) \citet{Warren:2007kw}, (51) \citet{Wright:2013bo}.}
    \end{tablenotes}
\end{table*}

\begin{table*}
    \centering
    \caption{Absolute AllWISE and PS1 Photometry for the Volume-Limited Sample}
    \label{tab:sample3}
    \begin{tabular}{lccccccc}
         \toprule
         \toprule
         \multicolumn{1}{c}{} & \multicolumn{3}{c}{AllWISE} & \multicolumn{4}{c}{PS1}\\
         \cmidrule(lr){2-4}\cmidrule(lr){5-8}
         Object & \Wone & \Wtwo & \Wthr & $r$  & $i$ & $z$ & $y$\\
          & (mag) & (mag) & (mag) & (mag) & (mag) & (mag) & (mag)\\
         \cmidrule(lr){1-8}
         SDSS J000013.54+255418.6 & $\nodata$ & $\nodata$ & $\nodata$ & $\nodata$ & $\nodata$ & 18.42 $\pm$ 0.06 & 16.67 $\pm$ 0.06 \\
         WISE J000517.48+373720.5 & 17.28 $\pm$ 0.10 & 13.81 $\pm$ 0.05 & 12.31 $\pm$ 0.24 & $\nodata$ & $\nodata$ & $\nodata$ & $\nodata$ \\
         PSO J003.3437-11.7168 & 13.53 $\pm$ 0.17 & 12.35 $\pm$ 0.18 & $\nodata$ & $\nodata$ & $\nodata$ & 18.46 $\pm$ 0.18 & 16.72 $\pm$ 0.17 \\
         2MASSW J0015447+351603 & 10.63 $\pm$ 0.02 & 10.38 $\pm$ 0.02 & 9.79 $\pm$ 0.10 & 19.22 $\pm$ 0.03 & 17.02 $\pm$ 0.02 & 15.65 $\pm$ 0.02 & 14.72 $\pm$ 0.02 \\
         PSO J004.6359+56.8370 & $\nodata$ & $\nodata$ & $\nodata$ & $\nodata$ & $\nodata$ & 18.90 $\pm$ 0.21 & 17.19 $\pm$ 0.18 \\
         PSO J007.9194+33.5961 & 12.05 $\pm$ 0.18 & 11.54 $\pm$ 0.18 & 10.49 $\pm$ 0.39 & $\nodata$ & 20.29 $\pm$ 0.20 & 17.88 $\pm$ 0.18 & 16.88 $\pm$ 0.18 \\
         2MASS J00320509+0219017 & 10.53 $\pm$ 0.03 & 10.29 $\pm$ 0.04 & 10.04 $\pm$ 0.40 & 18.95 $\pm$ 0.06 & 16.74 $\pm$ 0.02 & 15.37 $\pm$ 0.02 & 14.41 $\pm$ 0.03 \\
         ULAS J003402.77-005206.7 & 16.19 $\pm$ 0.14 & 13.72 $\pm$ 0.07 & $\nodata$ & $\nodata$ & $\nodata$ & $\nodata$ & $\nodata$ \\
         2MASS J00345157+0523050 & 15.49 $\pm$ 0.07 & 12.95 $\pm$ 0.06 & 12.18 $\pm$ 0.31 & $\nodata$ & $\nodata$ & 20.38 $\pm$ 0.06 & 18.50 $\pm$ 0.06 \\
         WISE J003829.05+275852.1 & 17.21 $\pm$ 0.16 & 14.12 $\pm$ 0.07 & 12.14 $\pm$ 0.34 & $\nodata$ & $\nodata$ & $\nodata$ & $\nodata$ \\
         HD 3651B & $\nodata$ & $\nodata$ & $\nodata$ & $\nodata$ & $\nodata$ & $\nodata$ & 17.99 $\pm$ 0.04 \\
         WISE J004024.88+090054.8 & 15.21 $\pm$ 0.08 & 13.05 $\pm$ 0.07 & $\nodata$ & $\nodata$ & $\nodata$ & 20.07 $\pm$ 0.09 & 18.13 $\pm$ 0.06 \\
         WISEPC J004928.48+044100.1 & 12.41 $\pm$ 0.11 & 11.94 $\pm$ 0.11 & $\nodata$ & $\nodata$ & 20.63 $\pm$ 0.19 & 17.98 $\pm$ 0.10 & 16.93 $\pm$ 0.10 \\
         WISE J004945.61+215120.0 & 16.65 $\pm$ 0.07 & 13.76 $\pm$ 0.05 & 12.31 $\pm$ 0.19 & $\nodata$ & $\nodata$ & 21.73 $\pm$ 0.05 & 20.09 $\pm$ 0.06 \\
         CFBDS J005910.90-011401.3 & 16.97 $\pm$ 0.13 & 13.80 $\pm$ 0.06 & $\nodata$ & $\nodata$ & $\nodata$ & $\nodata$ & $\nodata$ \\
         \bottomrule
    \end{tabular}
    \begin{tablenotes}
            \item \textbf{Notes.} Absolute AllWISE and PS1 photometry for all objects in the volume-limited sample (Table~\ref{tab:sample1}), calculated using the parallaxes in Table~\ref{tab:sample1}. The AllWISE photometry reference is \citet{Cutri:2014wx} for all objects except the T3~dwarf 2MASS~J02132062+3648506C \citep{Deacon:2017kd}. The PS1 photometry reference is \citet{Chambers2016} for all objects except the T8~dwarf 2MASSI~J0415195$-$093506 \citep{Best2018}. This table is available in its entirety in a machine-readable form in the online journal. A portion is shown here.
    \end{tablenotes}
\end{table*}

\subsection{Analytical Methods}
As an initial quality assessment and to identify outliers, we fit polynomials to the photometry as a function of spectral type for each band using weighted total least squares and orthogonal distance regression. Both methods produced visually reasonable fits with no major outliers in all bands. We then used the Monte Carlo approach of \citet{Dupuy2012}, re-sampling normally distributed absolute magnitude errors and uniformly distributed spectral type errors (between their quoted limits) to obtain $10^4$ re-sampled points for each object. This approach allows us to account for the uncertainties of both absolute magnitudes and spectral type, the latter assumed to be $\pm1$ unless indicated otherwise. With each data point randomly sampled about its respective uncertainties, we then performed a least-squares fit to all of the re-sampled points and obtained a final $n$th degree polynomial function for each band. We used the same degree for the MKO, 2MASS, and AllWISE polynomials as in \cite{Dupuy2012}: sixth degree polynomials for all MKO bands, {\JtwoM} and {\HtwoM}, and fourth degree for {\KtwoM} and all AllWISE bands. The PS1 bands had not been  fit previously; we used sixth degree polynomials for {\zPS} and {\yPS} bands, fourth degree for {\iPS} band, and second degree for {\rPS} band. For all bands, these were the lowest order polynomials that captured the major trends in the data. Higher order functions produced nominal changes in the RMS about fits, and led to over-fitting in some cases.

\subsection{Results}

\subsubsection{Absolute Magnitude vs. Spectral Type Relations for Single Field Objects}
\begin{figure*}
    \centering
    \subfloat{\includegraphics[scale=0.28]{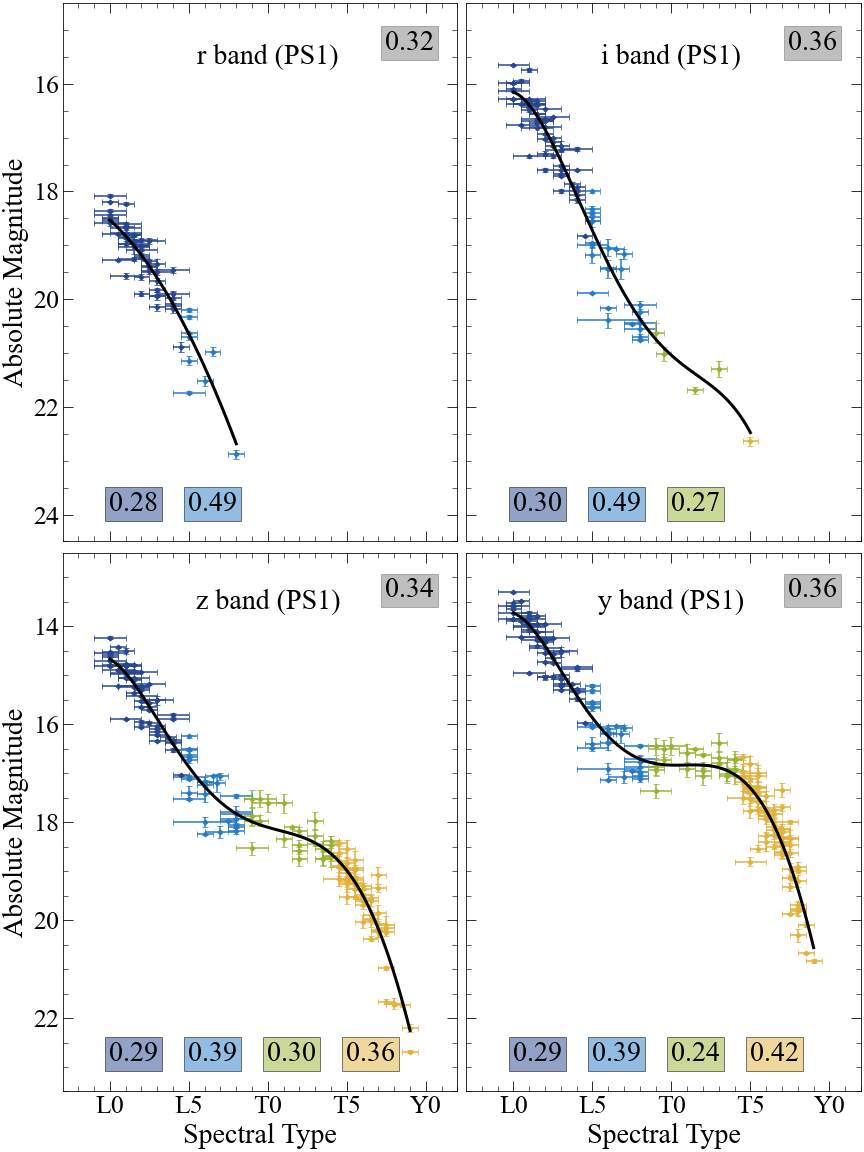}}\quad
    \subfloat{\includegraphics[scale=0.28]{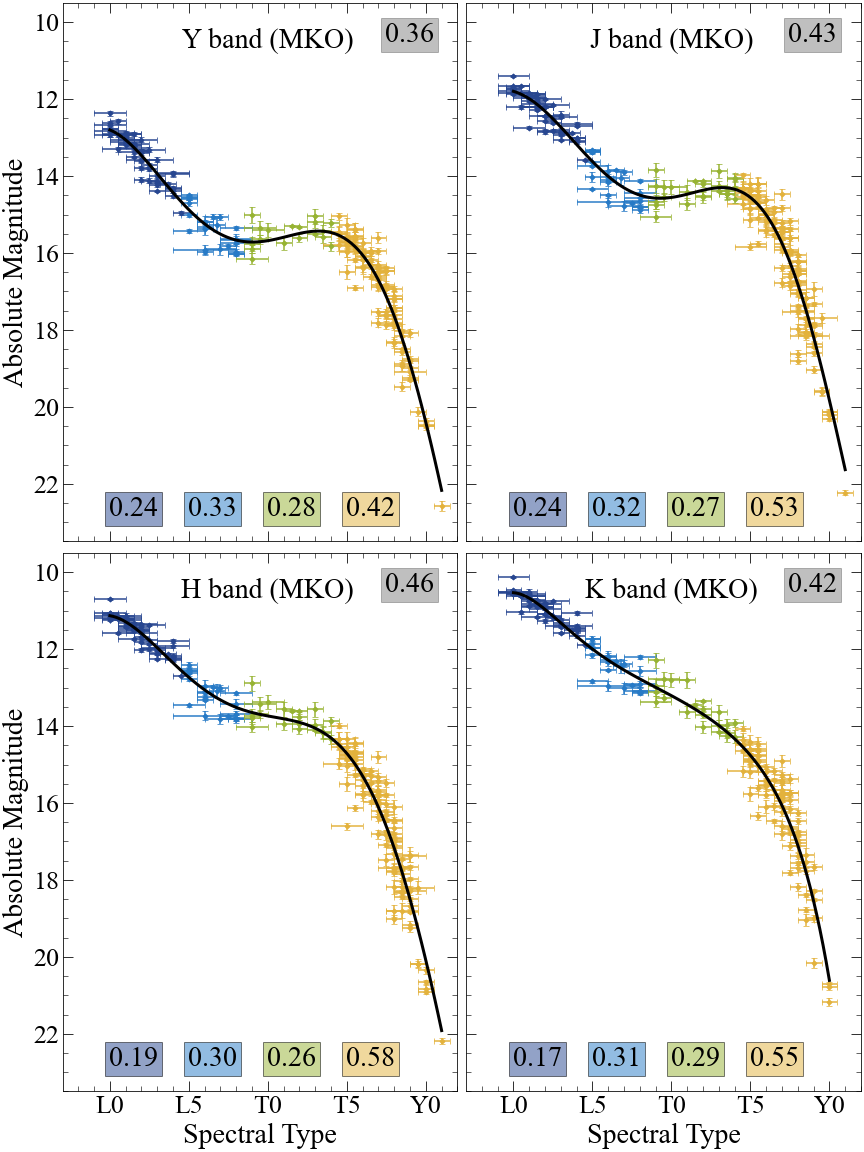}}
    \caption{Absolute magnitude as a function of spectral type for single ultracool field dwarfs with parallax measurements for $YJHK\textsubscript{MKO}$ and $rizy\textsubscript{P1}$ photometry. The solid black curves show the polynomial fits for our volume-limited sample (Table \ref{tab:polydata}). Polynomials are limited to the spectral type ranges for which objects were detected in each band. The total RMS about the fit is indicated in the gray box at upper right in each plot. The RMS about four spectral type ranges are given at the bottom: L0--L4.5 (dark blue), L5--L8.5 (light blue), L9--T4 (green), and $\geqslant$T4.5 (yellow). The polynomial functions display clean trends with no major outliers.}
    \label{fig:poly_mko-ps1}
\end{figure*}

\begin{figure*}
    \centering
    \subfloat{\includegraphics[scale=0.3]{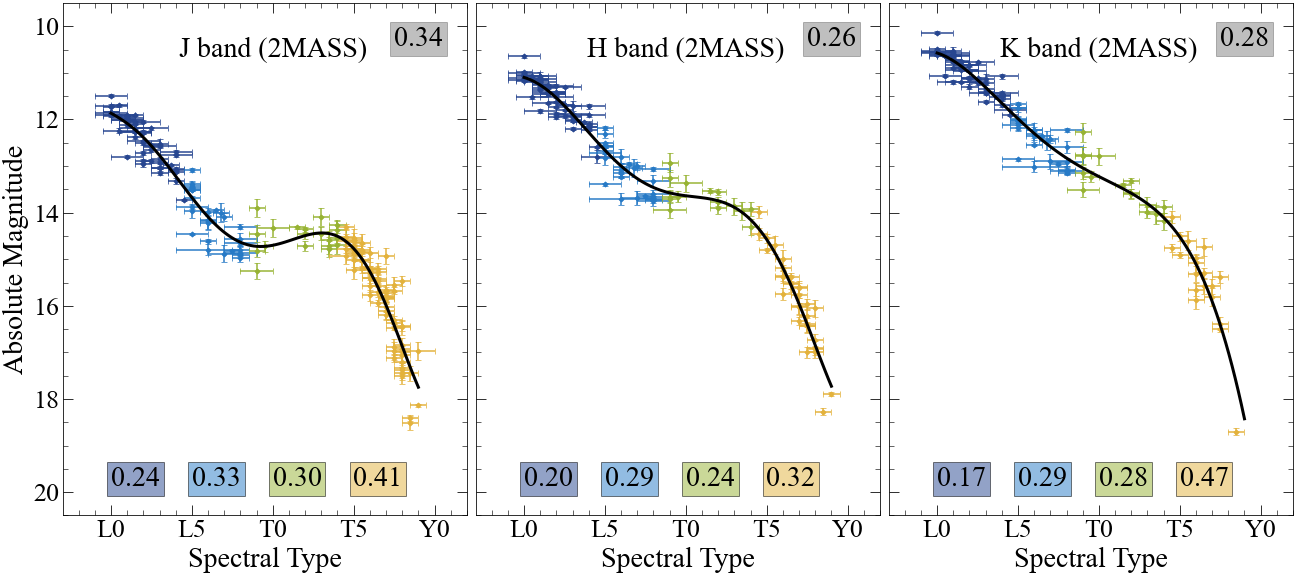}}\\
    \subfloat{\includegraphics[scale=0.3]{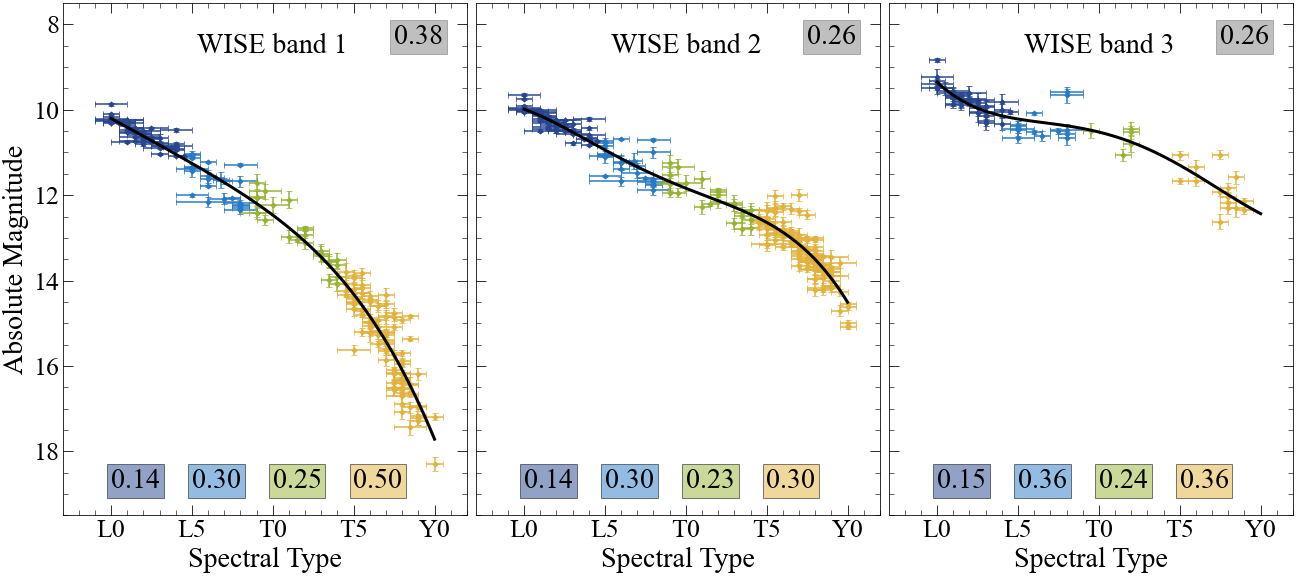}}
    \caption{Absolute magnitude as a function of spectral type (same as Figure~\ref{fig:poly_mko-ps1}) for 2MASS and AllWISE photometry.}
    \label{fig:poly_2mass-wise}
\end{figure*}

We present our polynomial fits in Table~\ref{tab:polydata}. For each polynomial, we include an RMS over all valid spectral types as well as RMS for up to four smaller spectral type ranges. In all bands, late-T dwarfs exhibit more scatter in absolute magnitude than earlier spectral types. Figures~\ref{fig:poly_mko-ps1} and \ref{fig:poly_2mass-wise} show our polynomial fits plotted over our cleaned sample of field objects (Section~\ref{sec:sample}). We note the phenomenon recognized as the ``$J$-band bump," in which early-T dwarfs appear brighter than late-L dwarfs \citep[e.g.,][]{Looper2008} in $J$-band. This apparent brightening is also seen in the {\YMKO} band, although less pronounced. In {\HMKO}, {\HtwoM}, {\zPS}, and {\yPS} bands there is a more gradual decline or plateau in absolute magnitude through the L/T transition. This prominent feature corresponds to the evolution of condensate clouds in the photospheres of late-L~dwarfs as they age and cool into T~dwarfs, whose photospheres are relatively cloud-free. One explanation for this is the ``patchy clouds'' model that suggests the disruption of clouds at $T_{\rm eff}\approx1200$~K allows for hotter flux from cloud-free regions to dominate the total emitted flux \citep{Ackerman2001, Burgasser2002}. Another hypothesis, the ``sudden downpour'' model, suggests that the rapid increase in efficiency of sedimentation causes the dusty clouds to condense out around $T_{\rm eff}\approx1300$ K \citep{Knapp2004a}.

\subsubsection{Young Objects, Subdwarfs, and Binaries}
\begin{figure*}
    \centering
    \subfloat{\includegraphics[scale=0.28]{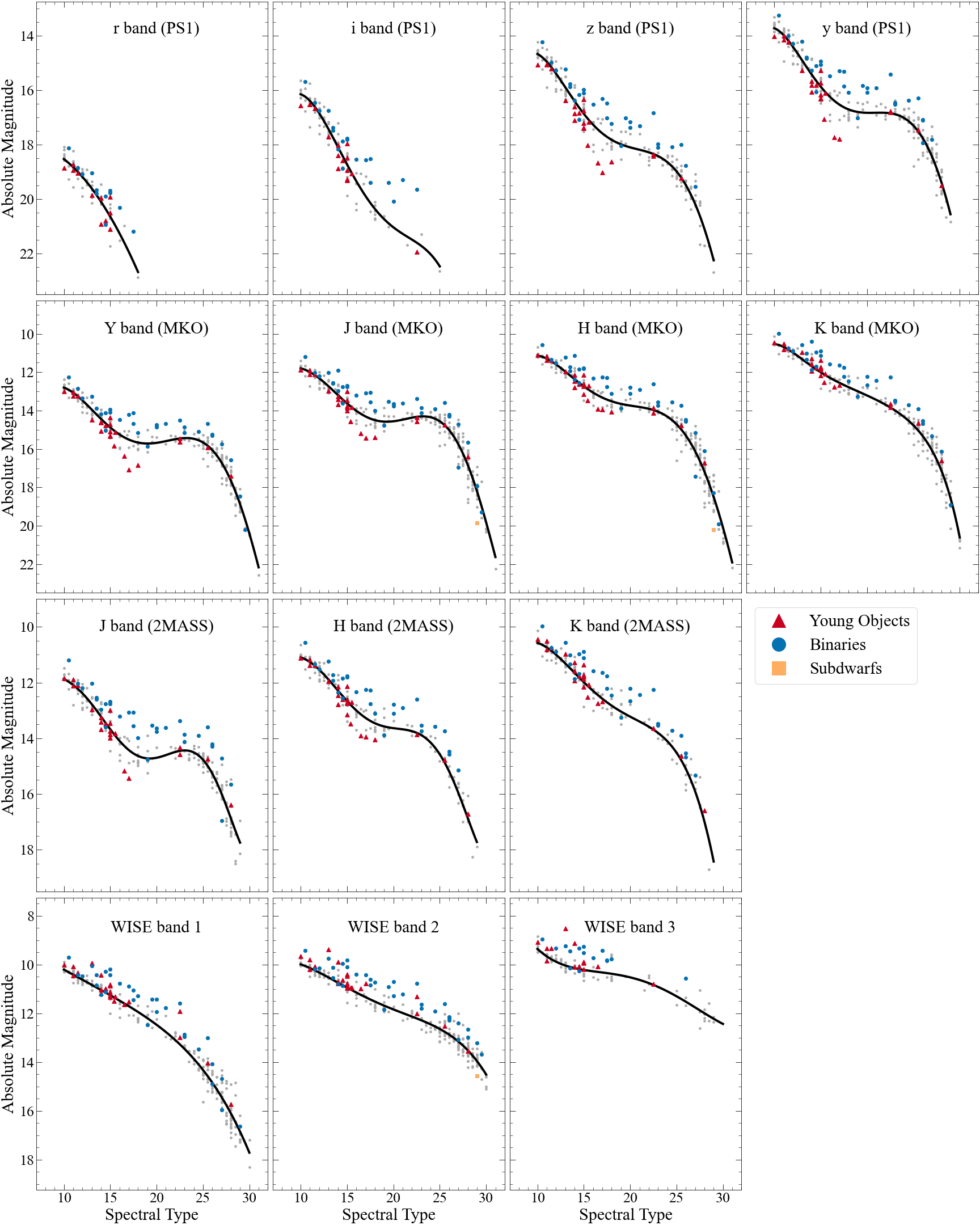}}
    \caption{Absolute magnitudes of young brown dwarfs (red triangles), known binaries (blue circles), and subdwarfs (yellow squares) compared to field objects (grey) with the polynomials from Figures~\ref{fig:poly_mko-ps1} and \ref{fig:poly_2mass-wise} over-plotted. Young objects appear fainter at $\approx$L4--L7 spectral types in 2MASS, MKO and PS1 ({\zPS} and {\yPS}) bands, and binaries appear significantly brighter than single objects.}
    \label{fig:yng-bi}
\end{figure*}

As described in Section~\ref{sec:sample}, young objects and older subdwarfs differ photometrically from their field-age counterparts. This is demonstrated in Figure~\ref{fig:yng-bi}, where young objects clearly deviate from the polynomial fits to the field objects in some bands. Early-L dwarf absolute magnitudes are comparable to the field sequence, but at mid-to-late L dwarfs, young objects are fainter in the MKO, 2MASS, and PS1 ({\zPS} and {\yPS}) bands \citep[e.g.,][]{Faherty2016a,LiuDupuy2016}. There are too few young or subdwarf T~dwarfs in our sample to assess possible systematic differences in absolute magnitudes. As discussed in \citet{Filippazzo2015}, \citet{LiuDupuy2016}, and \citet{Faherty2016a}, polynomial relations determined for field objects should not be used for young or suspected young sources. Additionally, known binaries appear overluminous compared with the single field group, as expected for unresolved sources that blend the light of two objects.

\begin{landscape}
\scriptsize

\begin{table}
\caption{Coefficients for Polynomial Fits of Absolute Magnitude to Spectral Type}\label{tab:polydata}
    \centering
    \begin{tabular}{lclllllllccccrr}
         \toprule
         \toprule
         \multicolumn{9}{c}{} & \multicolumn{5}{c}{RMS$^a$} \\
         \cmidrule(lr){10-14}
         $y$ & $x$ & \hspace{0.28in}$c_{0}$ & \hspace{0.28in}$c_{1}$ & \hspace{0.28in}$c_{2}$ & \hspace{0.28in}$c_{3}$ & \hspace{0.28in}$c_{4}$ & \hspace{0.28in}$c_{5}$ & \hspace{0.28in}$c_{6}$ & $r_1$ & $r_2$ & $r_3$ & $r_4$ & Total & Range$^b$\\
         \cmidrule(lr){1-15}
         \rPS & SpT & $\phantom{-}1.911\times10^{1}$ & $-3.775\times10^{-1}$ & $\phantom{-}3.201\times10^{-2}$
 & $\nodata$ & $\nodata$ & $\nodata$ & $\nodata$ & 0.28 & 0.49 & $\nodata$ & $\nodata$ & 0.32
 & L0--L6 \\
         \iPS & SpT & $\phantom{-}5.176\times10^{1}$ & $-9.916\times10^{0}$ & $\phantom{-}9.562\times10^{-1}$ & $-3.728\times10^{-2}$ & $\phantom{-}5.207\times10^{-4}$
 & $\nodata$ & $\nodata$ & 0.30 & 0.49 & 0.27 & $\nodata$ & 0.36
 & L0--T2 \\
         \zPS & SpT & $\phantom{-}5.642\times10^{1}$ & $-1.289\times10^{1}$ & $\phantom{-}1.452\times10^{0}$ & $-7.393\times10^{-2}$ & $\phantom{-}1.739\times10^{-3}$ & $-1.503\times10^{-5}$
 & $\nodata$ & 0.29 & 0.39 & 0.30 & 0.36 & 0.34
  & L0--T8 \\
         \yPS & SpT & $\phantom{-}6.358\times10^{1}$ & $-1.523\times10^{1}$ & $\phantom{-}1.707\times10^{0}$ & $-8.697\times10^{-2}$ & $\phantom{-}2.048\times10^{-3}$ & $-1.776\times10^{-5}$
 & $\nodata$ & 0.29 & 0.39 & 0.24 & 0.42 & 0.36
 & L0--T8 \\
         \YMKO & SpT & $\phantom{-}3.314\times10^{1}$ & $-4.581\times10^{0}$ & $\phantom{-}1.611\times10^{-1}$ & $\phantom{-}2.863\times10^{-2}$ & $-2.646\times10^{-3}$ & $\phantom{-}8.018\times10^{-5}$ & $-8.210\times10^{-7}$
 & 0.24 & 0.33 & 0.28 & 0.42 & 0.36
 & L0--T9 \\
         \JMKO & SpT & $\phantom{-}3.197\times10^{-2}$ & $\phantom{-}7.465\times10^{0}$ & $-1.630\times10^{0}$ & $\phantom{-}1.638\times10^{-1}$ & $-8.147\times10^{-3}$ & $\phantom{-}1.951\times10^{-4}$ & $-1.787\times10^{-6}$
 & 0.24 & 0.32 & 0.27 & 0.53 & 0.43
 & L0--T9 \\
         \HMKO & SpT & $\phantom{-}3.756\times10^{1}$ & $-7.055\times10^{0}$ & $\phantom{-}5.951\times10^{-1}$ & $-1.096\times10^{-2}$ & $-7.656\times10^{-4}$ & $\phantom{-}3.653\times10^{-5}$ & $-4.294\times10^{-7}$
 & 0.19 & 0.30 & 0.26 & 0.58 & 0.46
 & L0--T9 \\
         \KMKO & SpT & $\phantom{-}8.884\times10^{1}$ & $-2.728\times10^{1}$ & $\phantom{-}3.784\times10^{0}$ & $-2.701\times10^{-1}$ & $\phantom{-}1.065\times10^{-2}$ & $-2.216\times10^{-4}$ & $\phantom{-}1.915\times10^{-6}$
 & 0.17 & 0.31 & 0.29 & 0.55 & 0.42
 & L0--T9  \\
         \JtwoM & SpT & $-6.495\times10^{1}$ & $\phantom{-}3.158\times10^{1}$ & $-5.253\times10^{0}$ & $\phantom{-}4.463\times10^{-1}$ & $-2.020\times10^{-2}$ & $\phantom{-}4.629\times10^{-4}$ & $-4.208\times10^{-6}$
 & 0.24 & 0.33 & 0.30 & 0.41 & 0.34
 & L0--T8 \\
         \HtwoM & SpT & $-8.744\times10^{0}$ & $\phantom{-}1.042\times10^{1}$ & $-2.078\times10^{0}$ & $\phantom{-}2.010\times10^{-1}$ & $-9.965\times10^{-3}$ & $\phantom{-}2.439\times10^{-4}$ & $-2.329\times10^{-6}$
 & 0.20 & 0.29 & 0.24 & 0.32 & 0.26
 & L0--T8 \\
         \KtwoM & SpT & $\phantom{-}2.734\times10^{1}$ & $-4.828\times10^{0}$ & $\phantom{-}4.802\times10^{-1}$ & $-1.931\times10^{-2}$ & $\phantom{-}2.801\times10^{-4}$
 & $\nodata$ & $\nodata$ & 0.17 & 0.29 & 0.28 & 0.47 & 0.28
 & L0--T8 \\
         \Wone & SpT & $\phantom{-}8.812\times10^{0}$ & $-5.779\times10^{-3}$ & $\phantom{-}2.527\times10^{-2}$ & $-1.364\times10^{-3}$ & $\phantom{-}2.860\times10^{-5}$
 & $\nodata$ & $\nodata$ & 0.14 & 0.30 & 0.25 & 0.50 & 0.38
 & L0--T9 \\
         \Wtwo & SpT & $\phantom{-}1.405\times10^{1}$ & $-1.350\times10^{0}$ & $\phantom{-}1.429\times10^{-1}$ & $-5.643\times10^{-3}$ & $\phantom{-}7.994\times10^{-5}$
 & $\nodata$ & $\nodata$ & 0.14 & 0.30 & 0.23 & 0.30 & 0.26
 & L0--T9 \\
         \Wthr & SpT & $-3.904\times10^{0}$ & $\phantom{-}2.817\times10^{0}$ & $-2.088\times10^{-1}$ & $\phantom{-}6.729\times10^{-3}$ & $-7.652\times10^{-5}$
 & $\nodata$ & $\nodata$ & 0.15 & 0.36 & 0.24 & 0.36 & 0.26
 & L0--T9 \\
         \bottomrule
    \end{tabular}
    \begin{tablenotes}
            \item The coefficients are defined as $y=\sum_{i=0} c_ix^i$, where $y$ and $x$ are the quantities indicated in the first two columns. The numerical spectral types are defined such that $\text{L}0=10$ and $\text{Y}1=31$.
            \item $^a$ These columns give the RMS about the fit for four spectral types ranges: L0--L4 $(r_1)$, L4.5--L8 $(r_2)$, L8.5--T4 $(r_3)$, $\geqslant$T4.5 $(r_4)$; and for the total RMS about the fit over all spectral types. The {\rPS} and {\iPS} bands do not have RMS values for all ranges due to the lack of detected objects at later spectral types.
            \item $^b$ The spectral type ranges for which the fits are applicable. All ranges are inclusive.
        \end{tablenotes}
\end{table}

\setlength{\tabcolsep}{2.5pt}
\begin{table}
    \centering
    \caption{Summary of Constraints from {\TC} Models}
    \label{tab:TCresults}
    \begin{tabular}{lcccccccccccc}
         \toprule
         \toprule
         & {\SpT} (Num.) & {\Teff} (K) & {\logL} (dex) & {\JMKO} (mag) & {\HMKO} (mag) & {\KMKO} (mag) & {\JtwoM} (mag) & {\HtwoM} (mag) & {\KtwoM} (mag) & {\HtwoO-$J$} & {\HtwoO-$H$} & {\CHfour-$K$}\\
         Model & $\Delta \pm \sigma$ & $\Delta \pm \sigma$ & $\Delta \pm \sigma$ & $\Delta \pm \sigma$ & $\Delta \pm \sigma$ & $\Delta \pm \sigma$ & $\Delta \pm \sigma$ & $\Delta \pm \sigma$ & $\Delta \pm \sigma$ & $\Delta \pm \sigma$ & $\Delta \pm \sigma$ & $\Delta \pm \sigma$\\
         \cmidrule(lr){1-13}
         $M_1$ & $-0.23 \pm 0.91$ & $-131.0 \pm 221.4$ &  $-0.155 \pm 0.271$ & $\nodata$ & $\nodata$ & $\nodata$ & $\nodata$ & $\nodata$ & $\nodata$ & $\nodata$ & $\nodata$ & $\nodata$ \\
         $M_2$ & $-0.06 \pm 0.84$ & $\phantom{-}12.7 \pm 141.2$ & $\phantom{-}0.022 \pm 0.150$ & $-0.054 \pm 0.384$ & $-0.036 \pm 0.367$ & $-0.052 \pm 0.353$ & $-0.054 \pm 0.390$ & $-0.036 \pm 0.373$ & $-0.054 \pm 0.355$ & $\nodata$ & $\nodata$ & $\nodata$ \\
         $M_{3}$ & $\nodata$ & $\phantom{-}19.2 \pm 111.0$ &  $\phantom{-}0.022 \pm 0.120$ & $-0.062 \pm 0.307$ & $-0.057 \pm 0.293$ & $-0.065 \pm 0.292$ & $-0.056 \pm 0.307$ & $-0.053 \pm 0.293$ & $-0.057 \pm 0.290$ & $0.0 \pm 0.012$ & $0.0 \pm 0.009$ & $0.0 \pm 0.008$ \\ 
         $M_{3L}$ & $\nodata$ & $\phantom{-}20.2 \pm 117.3$ &  $\phantom{-}0.024 \pm 0.126$ & $-0.062 \pm 0.311$ & $-0.057 \pm 0.299$ & $-0.064 \pm 0.296$ & $-0.058 \pm 0.315$ & $-0.053 \pm 0.300$ & $-0.060 \pm 0.295$ & $0.0 \pm 0.012$ & $0.0 \pm 0.009$ & $0.0 \pm 0.007$ \\ 
         $M_{3T}$ & $\nodata$ & $13.1 \pm 69.3$ &  $\phantom{-}0.012 \pm 0.084$ & $-0.062 \pm 0.283$ & $-0.058 \pm 0.274$ & $-0.061 \pm 0.277$ & $-0.046 \pm 0.263$ & $-0.047 \pm 0.275$ & $-0.043 \pm 0.266$ & $0.0 \pm 0.011$ & $0.0 \pm 0.010$ & $0.0 \pm 0.011$ \\
         $M_{4L}$ & $\nodata$ & $\phantom{-}25.0 \pm 111.4$ &  $\phantom{-}0.029 \pm 0.122$ & $-0.073 \pm 0.295$ & $-0.065 \pm 0.284$ & $-0.072 \pm 0.284$ & $-0.070 \pm 0.295$ & $-0.065 \pm 0.282$ & $-0.069 \pm 0.284$ & $0.0 \pm 0.011$ & $0.0 \pm 0.009$ & $0.0 \pm 0.007$ \\ 
         $M_{4T}$ & $\nodata$ & $-3.0 \pm 32.8$ &  $-0.003 \pm 0.064$ & $\phantom{-}0.013 \pm 0.161$ & $\phantom{-}0.007 \pm 0.155$ & $-0.002 \pm 0.173$ & $\nodata$ & $\nodata$ & $\nodata$ & $0.0 \pm 0.006$ & $0.0 \pm 0.010$ & $0.0 \pm 0.013$ \\
         \bottomrule
    \end{tabular}
    \begin{tablenotes}
            \item \textbf{Notes.} This table lists the scatter ($\sigma$) and bias ($\Delta$) of the inferred labels compared with the reference labels for each \textit{Cannon} model described in Section~\ref{sec:TC_res}. The labels include numerical spectral type ({\SpT}), effective temperature ({\Teff}), bolometric luminosity ({\logL}), MKO photometry ({${JHK}_{\rm MKO}$}), 2MASS photometry ({${JHK}_{\rm 2M}$}), and {\SpT}-dependent spectral indices ({\HtwoO$_{JH}$} and {\CHfour$_{K}$}). In general, including photometry measurements and spectral indices as labels, as well as separating L and T dwarfs in the reference sets used to train {\TC} models, provided the best results. The scatter ($\sigma$) for ${JHK}_{\rm MKO}$ and ${JHK}_{\rm 2M}$ photometry for these models ($M_{3}$, $M_{3L}$, $M_{3T}$, $M_{4L}$, and $M_{4T}$) are smaller than the total RMS of our polynomial relations presented in Table~\ref{tab:polydata}.
            \item $M_1$ (Sec.~\ref{sec:simp_mod}) - Three-label model that includes {\Teff}, numerical {\SpT}, and {\logL} with L0--T8.5 dwarfs included in the reference set.
            \item $M_2$ (Sec.~\ref{sec:photo_mod}) - Nine-label model that includes {\Teff}, numerical {\SpT}, {\logL}, ${JHK}_{\rm MKO}$ and ${JHK}_{\rm 2M}$ photometry with L0--T8.5 dwarfs included in the reference set.
            \item $M_3$ (Sec.~\ref{sec:ind_mod}) - Eleven-label model that includes {\Teff}, {\logL}, ${JHK}_{\rm MKO}$ photometry, ${JHK}_{\rm 2M}$ photometry, and three $SpT$-dependent spectral indices with L0--T8.5 dwarfs included in the reference set. $M_{3L}$ includes only~L0--L8 dwarfs in the survey set. $M_{3T}$ includes only~L8--T8.5 dwarfs in the survey set.
            \item $M_{4L}$ (Sec.~\ref{sec:fin_mod}) - Eleven-label model that includes {\Teff}, {\logL}, ${JHK}_{\rm MKO}$ photometry, ${JHK}_{\rm 2M}$ photometry, and three $SpT$-dependent spectral indices with L0--L8 dwarfs included in the reference set.
            \item $M_{4T}$ (Sec.~\ref{sec:fin_mod}) - Eight-label model that includes {\Teff}, {\logL}, ${JHK}_{\rm MKO}$ photometry, and three $SpT$-dependent spectral indices with L8--T8.5 dwarfs included in the reference set.
    \end{tablenotes}
\end{table}

\end{landscape}

\subsubsection{Comparison to Previous Works}

\begin{figure*}
    \centering
    \includegraphics[scale=0.3]{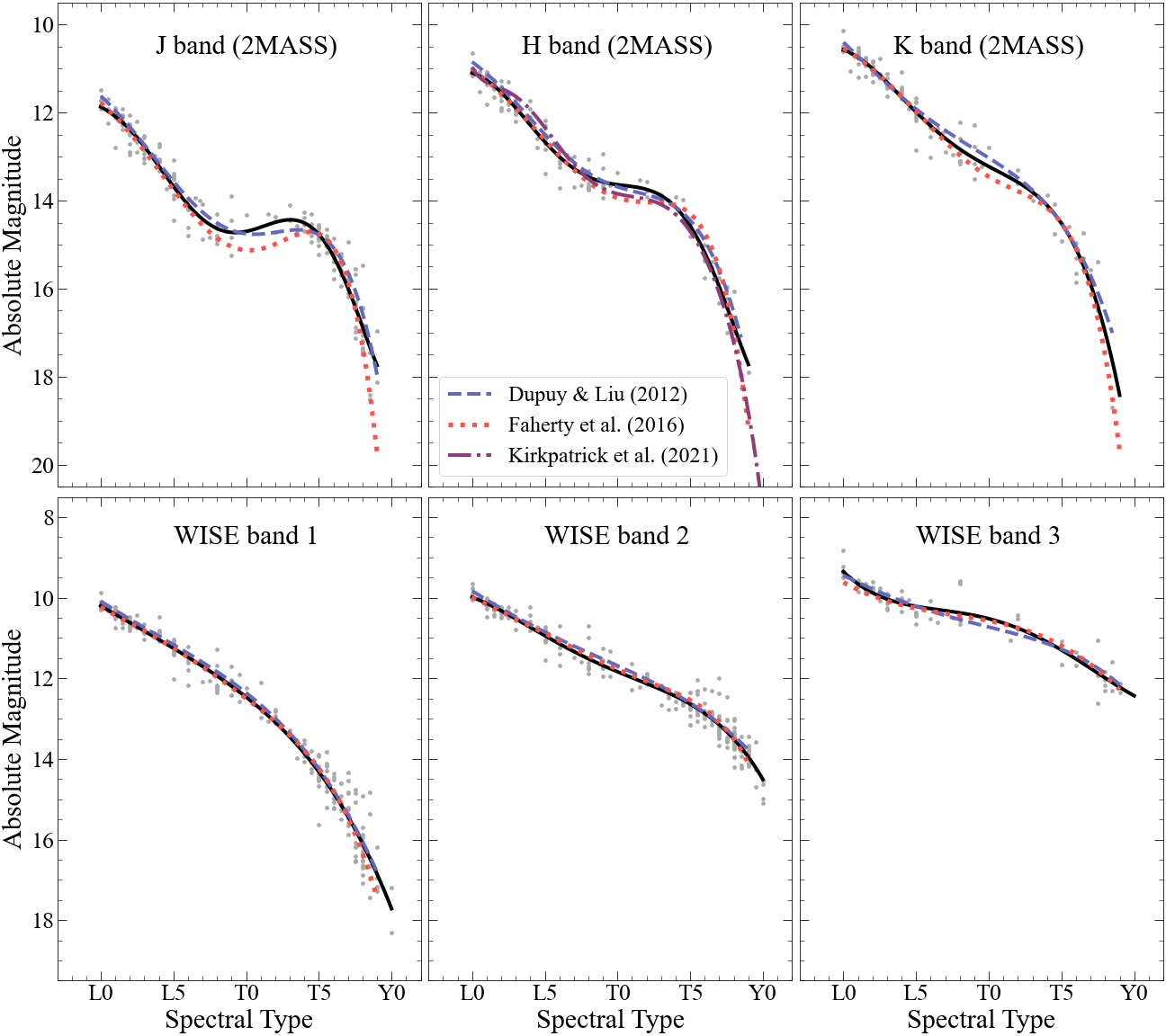}
    \caption{Comparison of our polynomial fits to previous work for 2MASS and AllWISE bands. The solid black curves show our fits (Table~\ref{tab:polydata}). Our polynomials are consistent with those of \citet[blue dashed lines]{Dupuy2012} and \citet[orange dotted lines]{Faherty2016a} for {\HtwoM}, {\KtwoM} and the AllWISE bands, and at early-L and late-T dwarfs for {\JtwoM}. For spectral types L8--T4, our {\JtwoM} fit has a sharper increase in flux, $\approx$0.5 mag brighter and occurring $\approx$1--2 spectral subtypes earlier than previous works. The $H$-band polynomial of \citet[combining 2MASS and MKO photometry; purple dot-dashed line]{Kirkpatrick2021} is  consistent with ours but deviates by $\lesssim$0.2~mag except for mid-T dwarfs.}
    \label{fig:2MWISE_comp}
\end{figure*}

\begin{figure}
    \centering
    \includegraphics[width=\columnwidth]{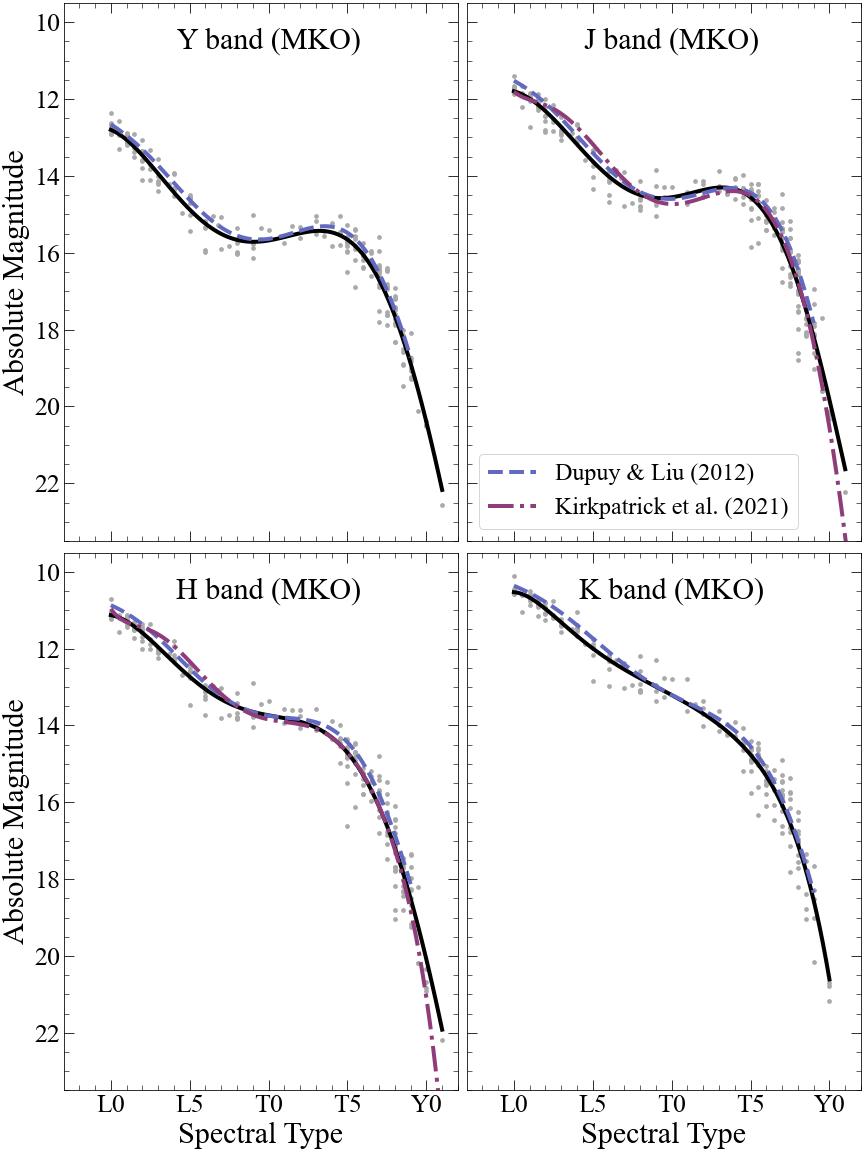}
    \caption{Same as Figure~\ref{fig:2MWISE_comp} for MKO bands. Our polynomial fits are consistent with those of \citet{Dupuy2012}; we note minor differences with those of \citet[$J$ and $H$ bands]{Kirkpatrick2021} for spectral types $\approx$L0--L8.} 
    \label{fig:MKO_comp}
\end{figure}

Our new polynomials, based on our volume-limited sample, should accurately represent the nearby brown dwarf population for spectral types L0--T8.5. 
Figures~\ref{fig:2MWISE_comp} and \ref{fig:MKO_comp} compare our polynomial relations to previously relations from \citet{Dupuy2012}, \citet{Faherty2016a}, and \citet{Kirkpatrick2021} for the MKO, 2MASS, and AllWISE bands. Since the fits of \citet{Dupuy2012} and \citet{Faherty2016a} do not use a volume-limited sample, our updated relations now provide a more robust standard for brown dwarf absolute magnitudes as a function of spectral type. The recent polynomials of \citet{Kirkpatrick2021} utilize a different volume-limited sample within 20 parsecs, but only for {\JMKO} and $H$ bands and mid-infrared \textit{Spitzer}/IRAC bands. \citet{Kirkpatrick2021} furthermore take $H$ band to be invariant across the MKO and 2MASS systems and combines them into a single polynomial, whereas we have determined independent fits for MKO and 2MASS $H$-band photometry.

For MKO bands, our fits are consistent with the relations of \citet{Dupuy2012} and \citet{Kirkpatrick2021}. We find no significant difference at the L/T transition, and any variation at L and late-T spectral types are within 0.3~mag of \citet{Dupuy2012}. The \citet{Kirkpatrick2021} {\JMKO} and $H$ relations appear $\approx$0.5~mag brighter than ours for spectral types L2--L8, possibly due to their inclusion of brighter mid-L outliers. Differences at late-T and early-Y spectral types for these bands may be accounted for by the presence of additional Y dwarfs in the all-sky volume-limited sample of \citet{Kirkpatrick2021}. In general, detection of late-T dwarfs and early-Y dwarfs is limited to $\lesssim20$~pc so our deeper, narrower sample contains fewer of these very faint objects. For 2MASS bands, we find our fits are consistent with previous work for early to mid-L (L0--L7) and mid to late-T~dwarfs (T5--T8). We note some differences at the L/T transition, especially in $J$-band. For spectral types L8--T4, our fit has a sharper increase in the $J$-band flux than \citet{Dupuy2012}, appears $\approx$0.5 mag brighter than \citet{Faherty2016a}, and occurs $\approx$1--2 spectral subtypes earlier than both. For AllWISE bands, our updated fits based on a volume-limited sample are very consistent with the polynomial relations of \citet{Dupuy2012} and \citet{Faherty2016a}.

\section{Machine Learning Models with {\TC}}
\label{sec:TC}

{\TC} is a data-driven approach designed to determine physical parameters, referred to as ``stellar labels'', of stars from spectroscopic data \citep{Ness2015, Ho2017a}. Using known labels (i.e., any measured quantity such as effective temperature, surface gravity, or elemental abundances) from a training set of stars, {\TC} develops a generative ``spectral model'' from their continuum-normalized spectra. The model assumes that stars with identical labels have similar spectra and the spectrum of each object is a smooth function of the labels that describe the object \citep{Ho2017a}. Models generated by {\TC} do not rely on physical models of the spectra, nor do they require accurate spectral types, which are heterogeneously determined and do not convey subtle differences in the spectral features that are unique to certain objects (e.g., peculiar objects). Furthermore, {\TC} can infer information directly from the spectra.

Previous applications of {\TC} have derived stellar labels from spectra in the context of large, high-resolution spectroscopic surveys like APOGEE and LAMOST \citep{Ness2015, Ness2016, Hogg2016, Ho2017a, Ho2017b}. It also has been shown to work efficiently and effectively on objects at lower signal-to-noise ($20\lesssim\text{S/N}\lesssim50$) with comparable accuracy to other physics-based approaches \citep{Ness2015}. Proven to work well with \textit{stellar} surveys, {\TC} has not been evaluated for \textit{substellar} objects like brown dwarfs. There are significant differences between the spectra of brown dwarfs and the stars used in previous applications of {\TC}; in particular, the prevalence of broad molecular absorption bands and (in most cases) the much lower resolution in brown dwarf spectra. Here we present our investigation into {\TC}'s potential to infer physical properties of brown dwarfs, its limitations, and future possibilities for its use. We explore the application of {\TC} and verify its methodology with low-resolution brown dwarf spectra, beginning with a simple model and successively adding to its complexity. Our main goal in this paper is to provide a proof of concept, demonstrating that physical properties can be inferred by {\TC} from low-resolution spectra containing broad molecular absorption bands. We do not examine all possible models or determine the most optimal model, but our results suggest how {\TC} might most successfully be implemented.

\subsection{Methodology}
The Cannon's procedures and requirements are described in detail in \citet{Ness2015}. Briefly, given a training set of spectra (hereinafter the \textit{reference set} with associated \textit{reference labels}), {\TC} performs a training step that generates the spectral model $g$ described by
\begin{equation}
    f_{n\lambda} = g(\bm{\ell_{n}}|\bm{\theta_{\lambda}}) + \text{noise}
	\label{eq:TCfunc}
\end{equation}
where $\bm{\theta_{\lambda}}$ is a coefficient vector, $\bm{\ell_{n}}$ is a label vector, and $f_{n\lambda}$ is the flux at each flux element. The noise takes into account the intrinsic scatter of the model and the uncertainty associated with the flux at each wavelength. In this training step, {\TC} utilizes the labels of the reference set to predict the flux at each wavelength and determine optimal coefficients for the spectral model. That model is then used to infer labels (known as \textit{survey labels}) from a set of survey objects (called the \textit{survey set}) in what is known as the test step. In this test step, {\TC} uses the spectral model, with coefficients $\bm{\theta_{\lambda}}$ determined in the training step, to optimize the labels of the survey set. In practice, the reference objects and survey objects will be different. For our investigation, the reference and survey sets were the same, allowing us to evaluate the effectiveness of {\TC} in analyzing the spectra of brown dwarfs. Additionally, the reference set used to train {\TC} should ideally be higher in S/N than the survey set, and it has been demonstrated that {\TC} can adequately transfer labels from high S/N reference objects to lower S/N survey objects \citep{Ness2015}.

To use {\TC}, the spectra need to meet the requirements outlined by \citet{Ness2015}. All spectra must come from the same telescope and reduction pipeline, and their labels must come from consistent sources. Any labels (i.e. any physical quantity of the object) can be included so long as they meet this condition. The spectra (for both reference and survey sets) must be normalized in a consistent manner that is independent of S/N, be radial velocity aligned, be sampled onto a common rest-frame wavelength grid, and have the same line-spread function. Finally, each flux value must have an associated error (or inverse variance).

\subsubsection{Preparing the Spectra}
\label{sec:prep}
The SpeX Prism Library \citep[SPL;][]{Burgasser2014} is a collection of over 2000 low-resolution ($\lambda/\Delta\lambda\approx$75--150) near-infrared spectra primarily of ultracool dwarfs acquired by the facility 0.8--5.5~$\mu$m SpeX spectrograph located on the 3.0~m NASA Infrared Telescope Facility (IRTF) \citep{Rayner2003}. As the largest library of L and T dwarf spectra currently available, the SPL has provided an organized and convenient resource for studying brown dwarfs. 

\begin{figure}
    \centering
    \includegraphics[width=\columnwidth, trim = 5mm 0 13mm 0, clip]{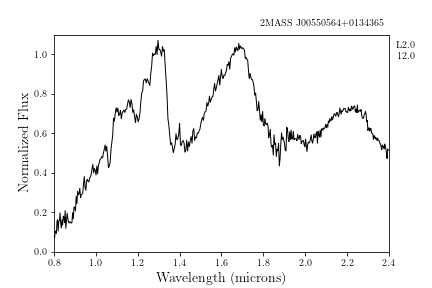}
    \caption{SpeX prism Spectrum ({$R$}$\sim$75) of the L2~dwarf 2MASS~J00550564+0134365 used in Section~\ref{sec:prep} for the wavelength calibration of the reference and survey sets. The flux has been normalized to the $J$-band peak at 1.27--1.28~$\mu$m. The spectrum illustrates the broad molecular absorption bands and lack of low-resolution continuum that epitomize brown dwarf spectra.}
    \label{fig:grid}
\end{figure}

We obtained over 1100 spectra of L and T dwarfs (L0--T9) from the SPL. We linearly interpolated each spectrum onto the same wavelength grid, for which we arbitrarily chose the wavelength grid of the {$R$}$\sim$75 spectrum of the L2 dwarf 2MASS~J00550564+0134365 \citep{Faherty2016a}. We restricted the wavelength range for our grid to 0.8--2.4 $\mu$m to exclude noisy regions at the extrema of the spectra. The wavelength grid contains 440~elements in a non-uniform spacing. {\TC} generally requires continuum-normalized spectra, but low-resolution L and especially T~dwarf spectra have little or no absorption-free continuum, so we opted to normalize each SPL spectrum to its $J$-band peak (1.27--1.28~$\mu$m), where absorption is minimized. As an example, we show the normalized spectrum of 2MASS~J00550564+0134365 in Figure~\ref{fig:grid}. 

\subsubsection{Determining Labels for Reference Objects}
\setlength{\tabcolsep}{3.75pt}

The reference labels used for this investigation include effective temperature ({\Teff}), numerical spectral type (\textit{SpT}, where L$0=10$ and T$8.5=28.5$), bolometric luminosity ({\logL}), MKO and 2MASS photometry (${JHK}_{\rm MKO}$ and ${JHK}_{\rm 2M}$) and the \textit{SpT}-dependent spectral indices of \citet{Burgasser2006}. To determine reference labels for {\Teff} and {\logL} in a uniform fashion, we used a single photometry-based polynomial relation from the literature for each label, along with the photometry compiled in Table~\ref{tab:sample1}.. The {\Teff} of each object in the SPL dataset was determined using the 2MASS $M_H$-based field-age polynomial of \citet{Filippazzo2015}. Similarly, bolometric luminosities were determined using the 2MASS $M_K$-based field-age polynomial of \citet{Dupuy2017}. We note that the quality of the inferred survey labels is limited by the accuracy and precision of the reference labels (quoted RMS of 29~K for {\Teff} and 0.05~dex for {\logL}).  Finally, the spectral indices of each object were measured using the SpeX Prism Library Analysis Toolkit\footnote{\url{http://pono.ucsd.edu/~adam/browndwarfs/splat/}}. Table~\ref{tab:labels} shows all reference labels used for {\TC} models described here. 

\begin{table*}
    \centering
    \caption{Reference Labels Used for {\TC} Models}
    \label{tab:labels}
    \begin{tabular}{lcccccccccccc}
         \toprule
         \toprule
         Object & {\SpT} & {\Teff} &  {\logL} & {\JMKO} & {\HMKO} & {\KMKO}& {\JtwoM} & {\HtwoM} & {\KtwoM} & {\HtwoO-$J$} & {\HtwoO-$H$} & {\CHfour-$K$}\\
          & (Num.) & (K) & (dex) & (mag) & (mag) & (mag) & (mag) & (mag) & (mag) & & & \\
         \cmidrule(lr){1-13}
         SDSS   J000013.54+255418.6 & 24.5 & 1283.5 & $-$4.803 & 13.98 & 13.99 & 14.07 & 14.31 & 13.98 & 14.09 & 0.325 & 0.373 & 0.259 \\
         2MASSI J0006205-172051 & 12.5 & 2030.1 & $-$3.818 & 12.61 & 11.73 & 11.00 & 12.67 & 11.65 & 11.02 & 0.828 & 0.780 & 1.068 \\
         2MASS J00100009-2031122 & 10 & 2260.9 & $-$3.630 & 11.83 & 11.16 & 10.59 & 11.87 & 11.10 & 10.61 & 1.014 & 0.904 & 1.025 \\
         2MASSI J0013578-223520 & 14 & 1737.4 & $-$4.175 & 13.53 & 12.51 & 11.85 & 13.60 & 12.42 & 11.87 & 0.761 & 0.725 & 0.994 \\
         2MASSW J0015447+351603 & 12 & 1998.5 & $-$3.860 & 12.58 & 11.80 & 11.09 & 12.72 & 11.73 & 11.10 & 0.866 & 0.839 & 1.042 \\
         \bottomrule
    \end{tabular}
    \begin{tablenotes}
            \item \textbf{Notes.} Reference labels used for all L0--T8.5 dwarfs: numerical spectral type ({\SpT}), effective temperature ({\Teff}), bolometric luminosity ({\logL}), MKO and 2MASS photometry ($JHK$ bands), and {\SpT}-dependent spectral indices ({\HtwoO$_J$}, {\HtwoO$_H$}, and {\CHfour$_K$}). The table is available in its entirety in a machine-readable form in the online journal. A portion is shown here.
    \end{tablenotes}
\end{table*}

\subsubsection{Identifying ``Continuum Pixels''}
\label{sec:pixels}

{\TC} requires both the reference and survey sets to be normalized as independently of S/N as possible. To do this, {\TC} performs a ``pseudo-continuum normalization'' process in which continuum pixels are defined as pixels whose normalized flux values are consistently close to 1 for all training spectra and vary minimally with changes in the label values. These pixels are identified using a running quantile of the spectra for user-specified wavelength ranges that are known to not be strongly affected by absorption features. Once these pixels are determined, {\TC} applies a continuum mask based on a fraction of the continuum pixels. 
For our data sets, we used label-dependent regions of the spectra (typically $J$, $H$, and $K$ bands) to generate our models, and we chose 4--9~per~cent of a running upper quantile (90th percentile) of the pixels in these regions as continuum pixels to be used for the continuum mask

We then fit the continuum using an N-th order sinusoid or Chebyshev function. We note that the performance of {\TC} greatly relies on the manner in which the spectra is \textit{Cannon}-normalized. We describe the specific choices we made for each model we tested in Section~\ref{sec:TC_res}. 

Since {\TC} relies on the continuum normalization of the set, and brown dwarf spectra have little to no absorption-free regions, it proved challenging to determine the most appropriate application of this pseudo-continuum normalization step for our data sets.

We learned that adjustments are necessary for an improved continuum mask and satisfactory fit, depending on the labels used and the objects in the reference set. 
We found the most appropriate method was typically to treat the $J$, $H$, and $K$ bands as three distinct spectral regions for {\TC} to normalize separately. Overall, it is important to specify wavelength ranges to which the labels are correlated. For example, if photometry measurements are used as labels, the wavelength ranges for these bands should be included in the continuum masking steps. We determined the best values for the parameters described here through repetitive trials, checking the normalization plots and identified continuum pixels in each case.

\subsection{Models and Results}
\label{sec:TC_res}

\subsubsection{$M_1$ Model: {\Teff}, {\logL}, and \SpT}
\label{sec:simp_mod}

\begin{figure*}
    \centering
    \subfloat[Reference Labels]{\includegraphics[scale=0.4]{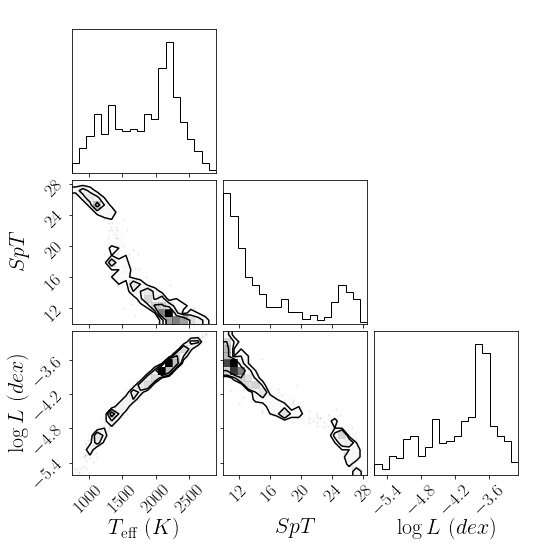}}\quad
    \subfloat[Survey Labels]{\includegraphics[scale=0.4]{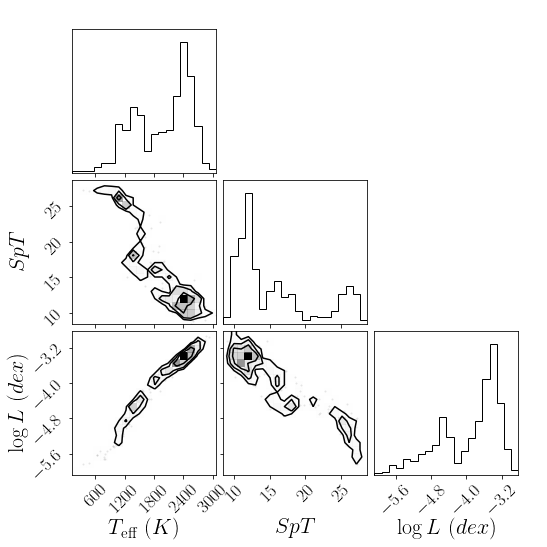}}
    \caption{Diagnostic plots of the reference (i.e., previously known) and survey (i.e., {\it Cannon}-determined) labels for a three-label {\it Cannon} model (Section~\ref{sec:simp_mod}) using effective temperature ({\Teff}), numerical spectral type (\SpT), and bolometric luminosity ({\logL}). Numerical spectral types are defined such that L$0=10$ and T$8.5=28.5$. 
    The histograms represent the distribution of the reference and survey labels of our sample. The contour plots show the density of labels within the label space.
    The distributions and correlations of the labels in the reference set are clearly reflected in the survey set, indicating that {\TC} can accurately infer information from the spectra of brown dwarfs.}
    \label{fig:diagnostics1}
\end{figure*}

We first constructed a simple \textit{Cannon} model using only three labels: {\Teff}, {\SpT}, and {\logL}. For the reference set, we used L0--T8.5 dwarfs with $9.4 \leq M_H \leq 18.1$ mag (2MASS) and $8.8 \leq M_K \leq 16.6$ mag (2MASS); these are the valid ranges for the polynomials of \citet{Filippazzo2015} and \citet{Dupuy2017} that we used to determine the {\Teff} and {\logL} reference labels, respectively. We included only objects with parallax uncertainty $\leq$20~per~cent for a final set of 390~L and T~dwarfs. We used this same set for the survey set as well, to evaluate the quality of the label transfer. Figure~\ref{fig:diagnostics1} shows diagnostic plots for this set comparing each label. We note the expected trends between {\Teff}, {\SpT}, and {\logL}: decreasing effective temperature and bolometric luminosity with increasing spectral type (early-L to late-T dwarfs) and the strong positive correlation of {\Teff} and {\logL}. To identify the continuum pixels we used a 90th percentile cutoff across three wavelength regions: 1.10--1.35~$\mu$m, 1.50--1.80~$\mu$m, and 2.0--2.35~$\mu$m. The ranges contain important label-dependent information in the $J$, $H$, and $K$ bands while largely avoiding the {\HtwoO} absorption features at $\sim$1.4~$\mu$m and $\sim$1.9~$\mu$m. 6~per~cent of the pixels were identified for the continuum mask, which we fit with a second-order sinusoid function. 

\begin{figure*}
    \centering
    \includegraphics[scale=0.33]{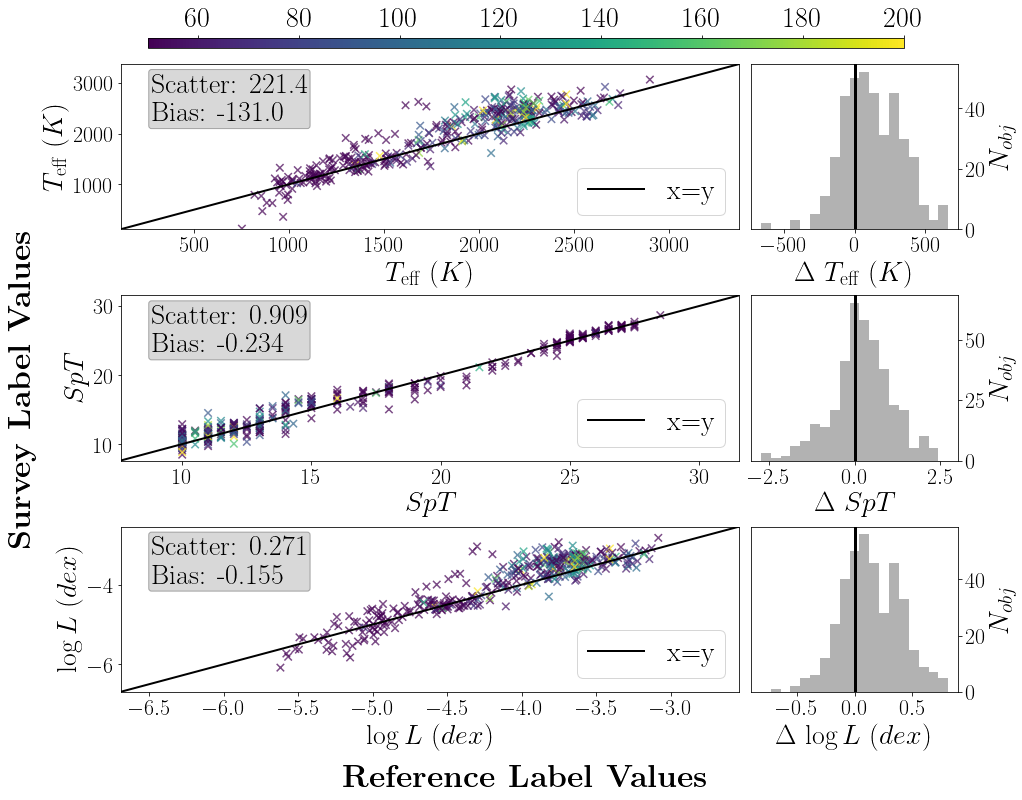}
    \caption{Diagnostic plots for the three-label model $M_1$ (Section~\ref{sec:simp_mod}) shown in  Figure~\ref{fig:diagnostics1}, comparing the reference labels (horizontal axis) to the \textit{Cannon}-generated survey labels (vertical axis). Each object is colour-coded by its spectroscopic S/N (colour bar at top). 
    The histograms show the number of objects $N_{\rm obj}$ in each label bin.
    There are clear correlations between the known and inferred values (albeit with significant bias and scatter in this simple model), showing that {\TC} can determine physical properties from the spectra of brown dwarfs.}
    \label{fig:Lab1}
\end{figure*}

Figure~\ref{fig:diagnostics1} can be used to visually assess the performance of the model. Comparing these diagnostic plots for the reference and survey sets, we see that the correlations between the reference labels are reproduced well in the inferred survey labels. Figure~\ref{fig:Lab1} compares the reference labels to the inferred survey labels for each object. The quality of the label transfer can be assessed by the scatter (i.e., RMS) and bias (i.e., offset) with respect to the equality (i.e., $x=y$) line. These values are provided in Table~\ref{tab:TCresults} for each model. In general, {\TC} infers greater values (i.e., higher {\Teff} and {\logL}, later spectral type) for the survey labels than the corresponding reference values for this $M_1$ model. We can compare these to the systematic uncertainties of the reference labels: 29 K for {\Teff} \citep{Filippazzo2015}, 0.05 dex for {\logL} \citep{Dupuy2017}, and (typically) 1 for spectral type classification. The uncertainty of the {\SpT} label in {\TC} models is comparable to that of other spectral classification methods \citep[e.g.,][]{Burgasser2010,Allers2013}, indicating that this model could be used as a tool to constrain the spectral types of brown dwarfs across broad surveys. (We note that specifying a single broad wavelength range across 1.1--2.3 $\mu$m for continuum normalization decreases the scatter of the label transfer for {\Teff} and {\logL}, but increases it for {\SpT} by $\sim0.3$.) Overall, this model infers more accurate labels for T dwarfs than L dwarfs. This likely arises from the dramatic changes in spectral morphology that occur across types T0--T8 \citep[e.g.,][]{Kirkpatrick:2005cv}, allowing {\TC} to more easily detect prominent features such as the shape of flux peaks and band strengths and determine more accurate label values, even at lower spectral resolution. 

\subsubsection{$M_2$ Model: Including Photometry as Labels}
\label{sec:photo_mod}
We then constructed a second \textit{Cannon} model with the same {\Teff}, {\logL}, and {\SpT} labels but also including MKO and 2MASS photometry in $J$, $H$, and $K$ bands. We used reference spectra from the same initial set of L and T dwarfs as in Section~\ref{sec:simp_mod}, this time extracting objects with absolute magnitude errors $<$0.2~mag in each band and parallax uncertainties $\leq$20~per~cent for a final set of 339~objects. However, when determining the wavelength ranges about which {\TC} creates the continuum mask, we found that specifying one wavelength range from 1.1--2.35 $\mu$m provided the best spectral model with the lowest scatter and bias for the inferred labels. This suggests that excluding {\HtwoO} features in the pseudo-continuum normalization process may remove informative pixels whose flux correlates with the $J$, $H$, or $K$ photometry labels, or that the overall spectral slope similarly correlates with the photometry. We identified the continuum mask using 4~per~cent of the flux and fit this with a second-order sinusoid function.

\begin{figure*}
    \centering
    \includegraphics[scale=0.25]{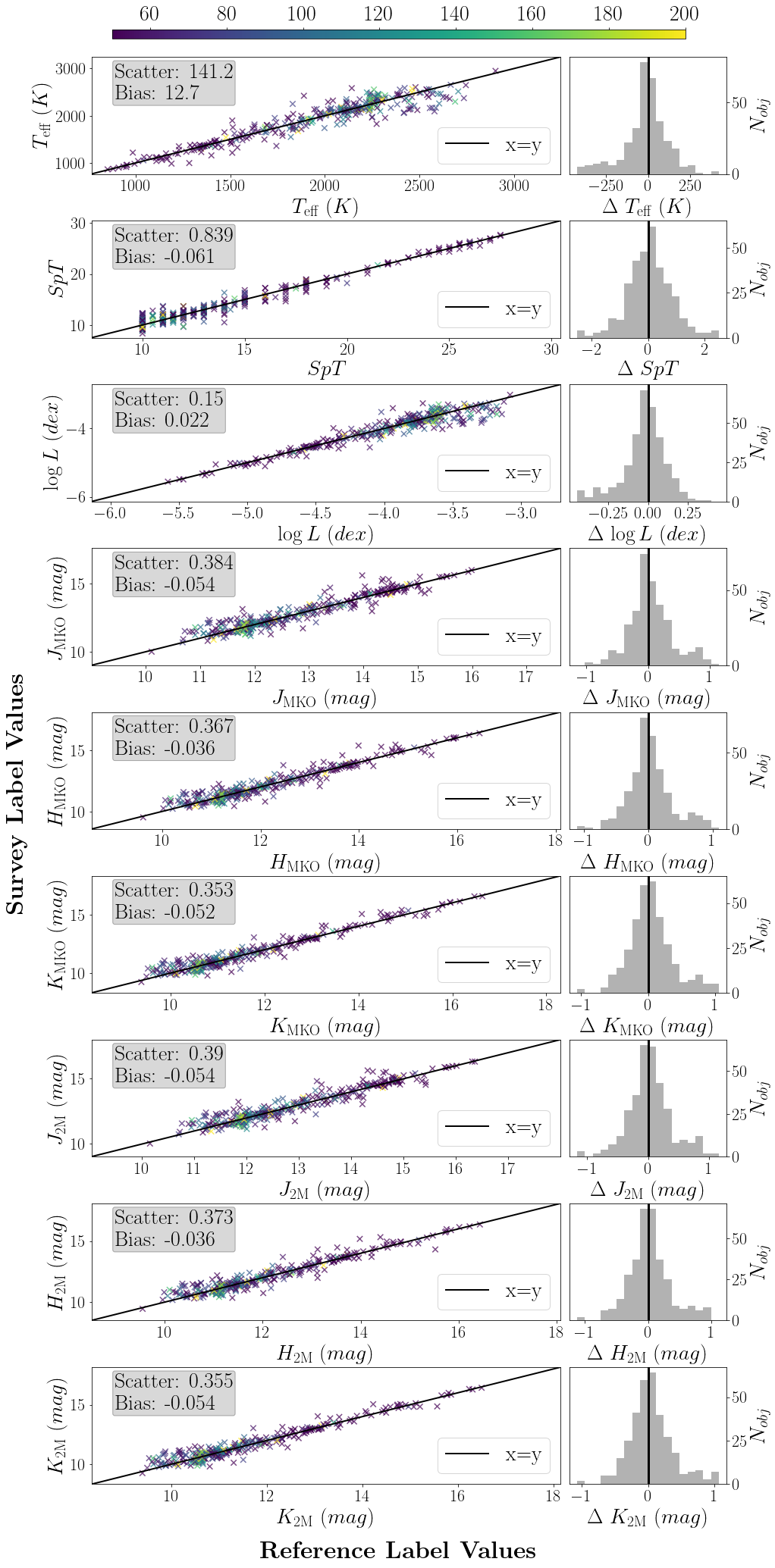}
    \caption{Same as Figure~\ref{fig:Lab1} but comparing the reference and survey values for {\TC} model $M_2$ (Section~\ref{sec:photo_mod}) including effective temperature ({\Teff}), numerical spectral type (\SpT), bolometric luminosity ({\logL}), and absolute magnitudes ($JHK\textsubscript{MKO}$ and $JHK\textsubscript{2M}$). Including photometry labels generates an improved \textit{Cannon} model with lower scatter and bias for the {\Teff}, {\SpT}, and {\logL} survey labels than those from the $M_1$ model (Figure~\ref{fig:Lab1}).}
    \label{fig:Lab2}
\end{figure*}

Figure~\ref{fig:Lab2} shows the reference and inferred labels with the scatter and bias of each, and we also present these values in Table~\ref{tab:TCresults}. We again find tighter agreement between the reference and inferred values for T dwarfs. The inclusion of photometry labels and the modification to the continuum mask creates an improved fit for the {\Teff}, {\SpT}, and {\logL} labels (36 per cent, 7.7 per cent, and 45 per cent decrease in scatter, respectively, relative the the $M_1$ model). The scatter of the inferred values for MKO photometry are comparable to the total RMS of the polynomial relations determined in Section~\ref{sec:poly}.

\subsubsection{$M_{3}$ Models: Including Spectral Indices as Labels}
\label{sec:ind_mod}
Spectral types, while useful labels for stellar and substellar classification, may not be as useful for {\TC} to model and infer from the spectra of brown dwarfs. {\TC} was designed to work with labels that vary continuously with changes in flux in the spectroscopic pixels, whereas spectral types are typically discrete values determined by visually comparing spectroscopic features to a spectral standard. An alternative classification method utilizes spectral indices which are readily calculated from the flux in specified wavelength ranges and can identify differences between spectra that track with spectral types but using continuous variables \citep[e.g.,][]{Burgasser2006}. Such indices can also indicate binarity and unusual surface gravity \citep{Allers2007, Allers2013, Burgasser2010, BardalezGagliuffi2014}.

We therefore constructed a new Cannon model replacing spectral types with spectral indices. We extracted 338~L and T~dwarfs, using the same absolute magnitude and parallax uncertainty cutoffs as in the $M_2$ model but now replacing spectral types with three {\SpT}-sensitive indices for labels: $H_2O_J$, $H_2O_H$, and ${CH_4}_K$ \citep{Burgasser2006}. We used 0.90--1.40 $\mu$m, 1.50--1.79 $\mu$m, and 1.99--2.35 $\mu$m as wavelength regions for the continuum mask, utilizing 6~per~cent of all pixels and a first-order sinusoid function.

\begin{figure*}
    \centering
    \includegraphics[scale=0.245]{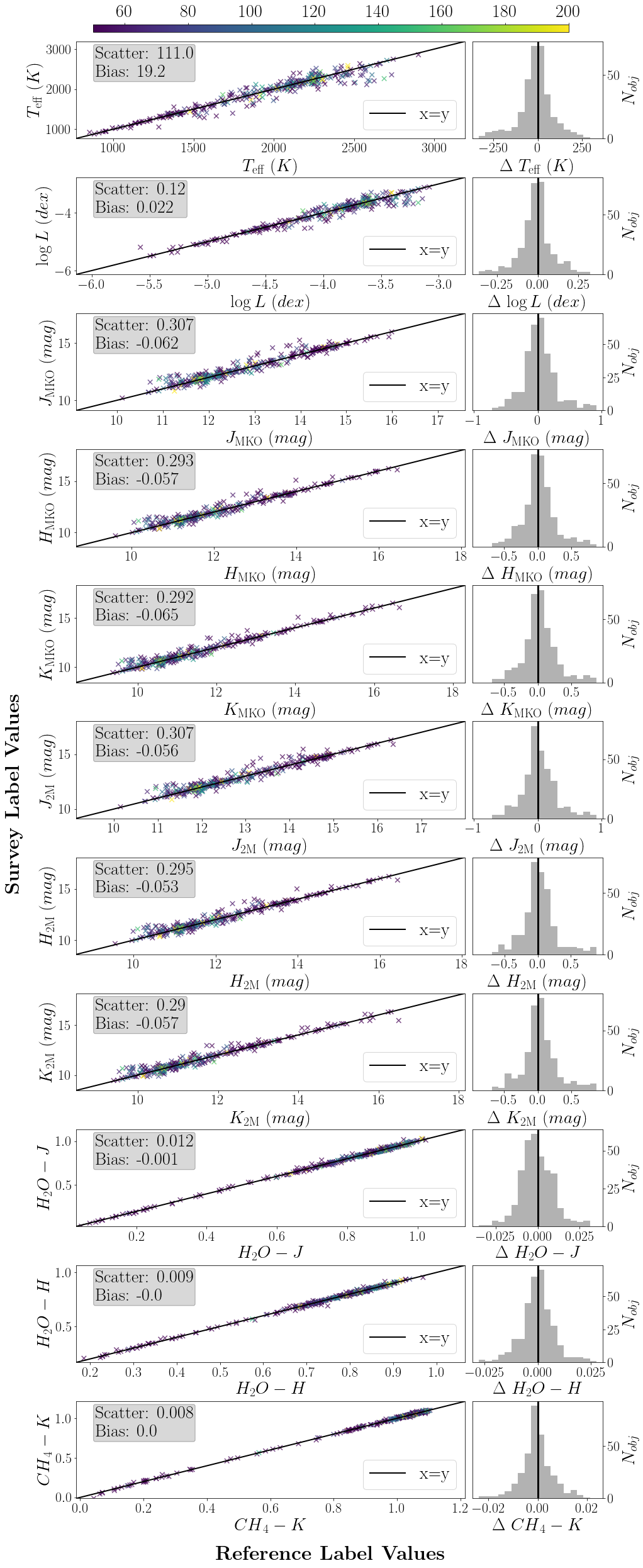}
    \caption{Same as Figures~\ref{fig:Lab1} and \ref{fig:Lab2} but comparing the reference and survey values for {\TC} model $M_3$ (Section~\ref{sec:ind_mod}) including effective temperature ({\Teff}), bolometric luminosity ({\logL}), absolute magnitudes (${JHK}_{MKO}$ and ${JHK}_{2M}$), and $H_2O_J$, $H_2O_H$, and ${CH_4}_K$ spectral indices for the L-dwarf (left) and T-dwarf (right) models. Including spectral indices generates an improved \textit{Cannon} model with lower scatter and bias for all inferred labels that were also generated in $M_2$ (Figure~\ref{fig:Lab2}).}
    \label{fig:Lab3a}
\end{figure*}

\begin{figure*}
    \centering
    \subfloat[{\TC} Model $M_{3L}$ for L0--L8 dwarfs]{\includegraphics[scale=0.245]{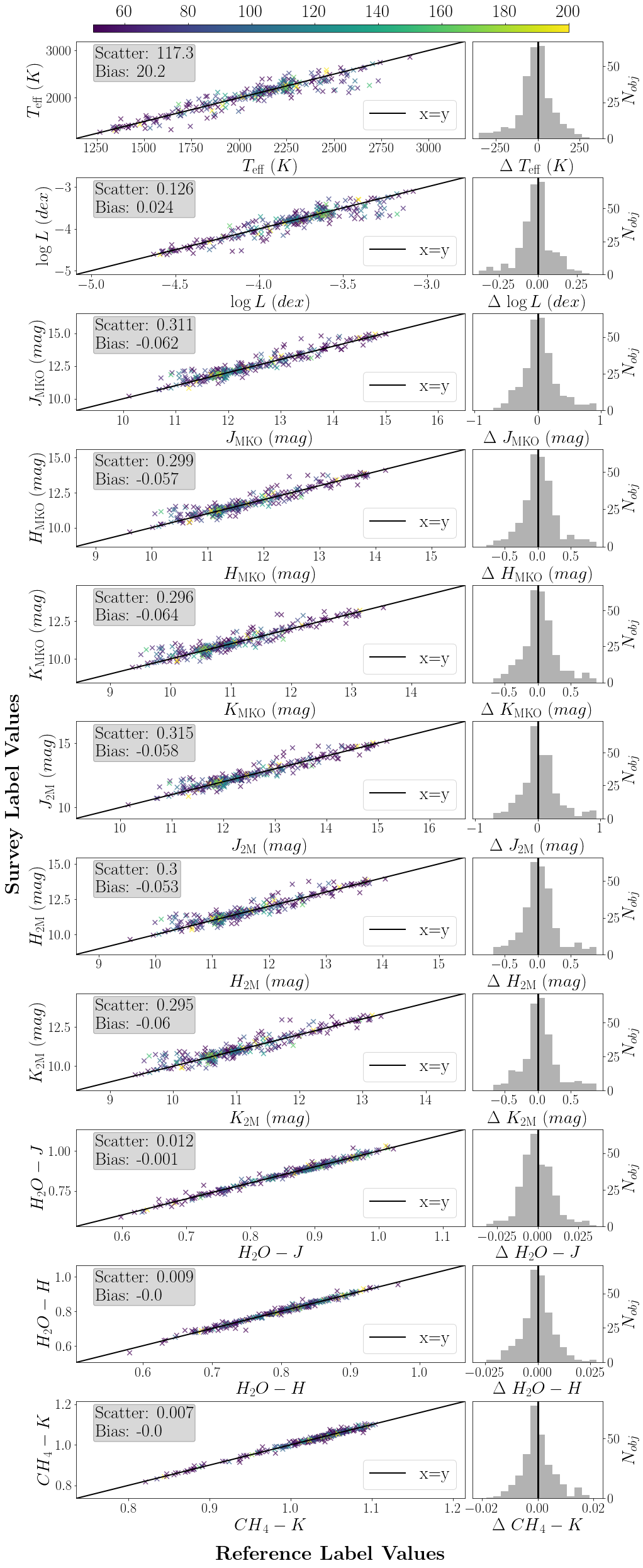}}\quad
    \subfloat[{\TC} Model $M_{3T}$ for L8--T8.5 dwarfs]{\includegraphics[scale=0.245]{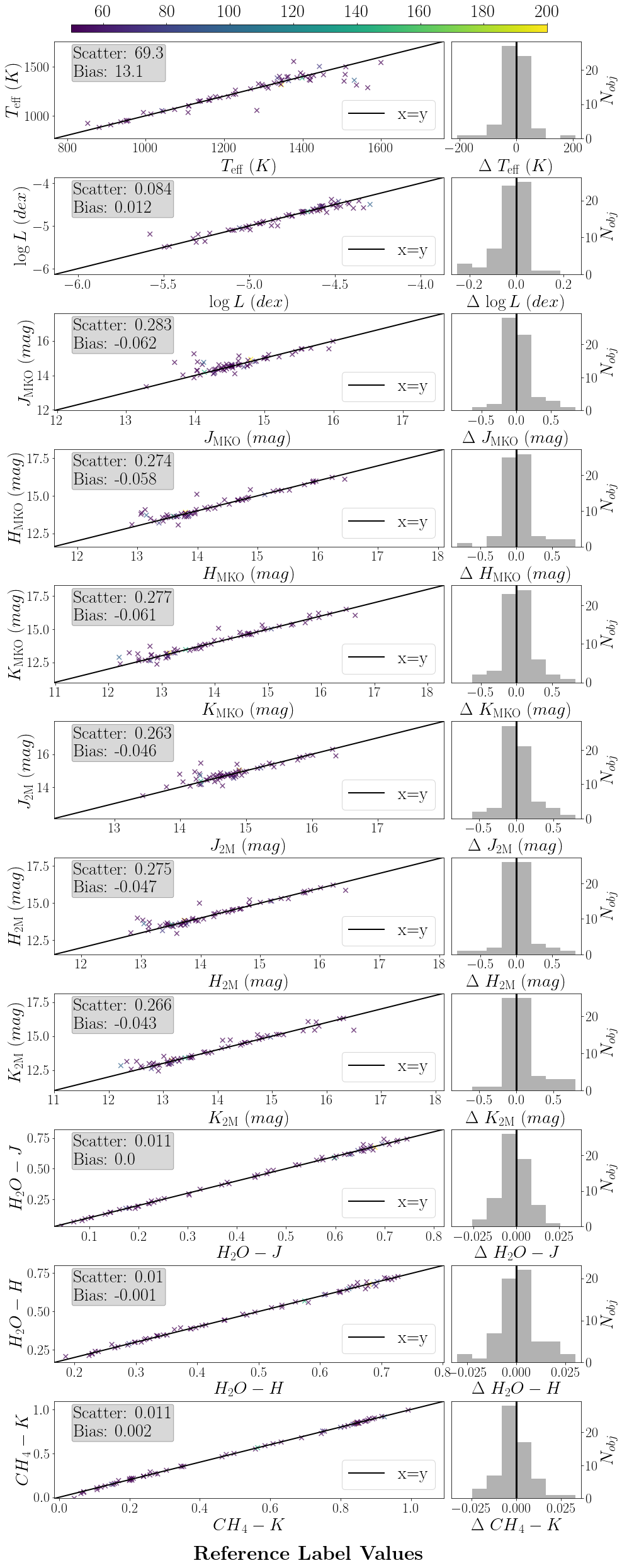}}
    \caption{Same as Figure~\ref{fig:Lab3a} but comparing the reference and survey values for distinct survey sets containing L0--L8 (Model $M_{3L}$, left) and L8--T8.5 (Model $M_{3T}$, right) dwarfs. These distinct survey sets show that {\TC} can infer properties of T~dwarfs with lower scatter and bias than for L~dwarfs.}
    \label{fig:Lab3b}
\end{figure*}

In addition to using the full reference set as the survey set (model $M_3$; Figure~\ref{fig:Lab3a}), we also ran this model with survey sets of L0--L8~dwarfs and L8--T8.5~dwarfs (models $M_{3L}$ and $M_{3T}$; Figure~\ref{fig:Lab3b}), in order to separately evaluate the model's performance on L and T dwarfs. The scatter and bias for each label are provided in Table~\ref{tab:TCresults} for all three models. We note a decrease in the scatter and bias for all labels with respect to the previous models that used numerical {\SpT} as labels, indicating that {\TC} infers labels with higher fidelity when given spectral indices rather than spectral types as labels. The inferred $JHK_{\rm MKO}$ for both the $M_{3L}$ and $M_{3T}$ survey sets have a lower RMS than the total RMS of our polynomial fits (Section \ref{sec:poly}). It is clear that {\TC} determines more accurate labels for T dwarfs than L dwarfs, as the scatter for the L dwarf survey set in each label is two or three times greater than for the corresponding label in the T dwarf survey set.


\subsubsection{$M_{4}$ Models: Separating L and T Dwarfs in the Reference Set}
\label{sec:fin_mod}
We constructed a final pair of models by separating L and T~dwarfs in the reference set. For the L-dwarf model, $M_{4L}$, we extracted 287 objects from the previous $M_3$ model set with spectral types L0--L8 (inclusive). We used the same labels as for the $M_3$ model. We used 0.90--1.10$\mu$m, 1.10--1.4 $\mu$m, 1.50--1.79 $\mu$m, and 1.99--2.35 $\mu$m as wavelength regions for the continuum mask, now identifying $Y$ band as a distinct region. The T-dwarf model, $M_{4T}$, contained 109 objects with spectral types L8--T8.5 (inclusive), including the L8 and L9 dwarfs at the warm end of the L/T transition. Due to the limited number of T~dwarfs with precise 2MASS photometry measurements, we excluded $JHK_{\rm 2M}$ as labels for this $M_{4T}$ model. We used 0.90--1.15$\mu$m, 1.15--1.4 $\mu$m, 1.50--1.79 $\mu$m, and 1.99--2.35 $\mu$m as wavelength regions for the continuum mask. We again utilized 6~per~cent of all pixels and a first-order sinusoid function.

\begin{figure*}
    \centering
    \subfloat[{\TC} Model $M_{4L}$ for L0--L8 dwarfs]{\includegraphics[scale=0.245]{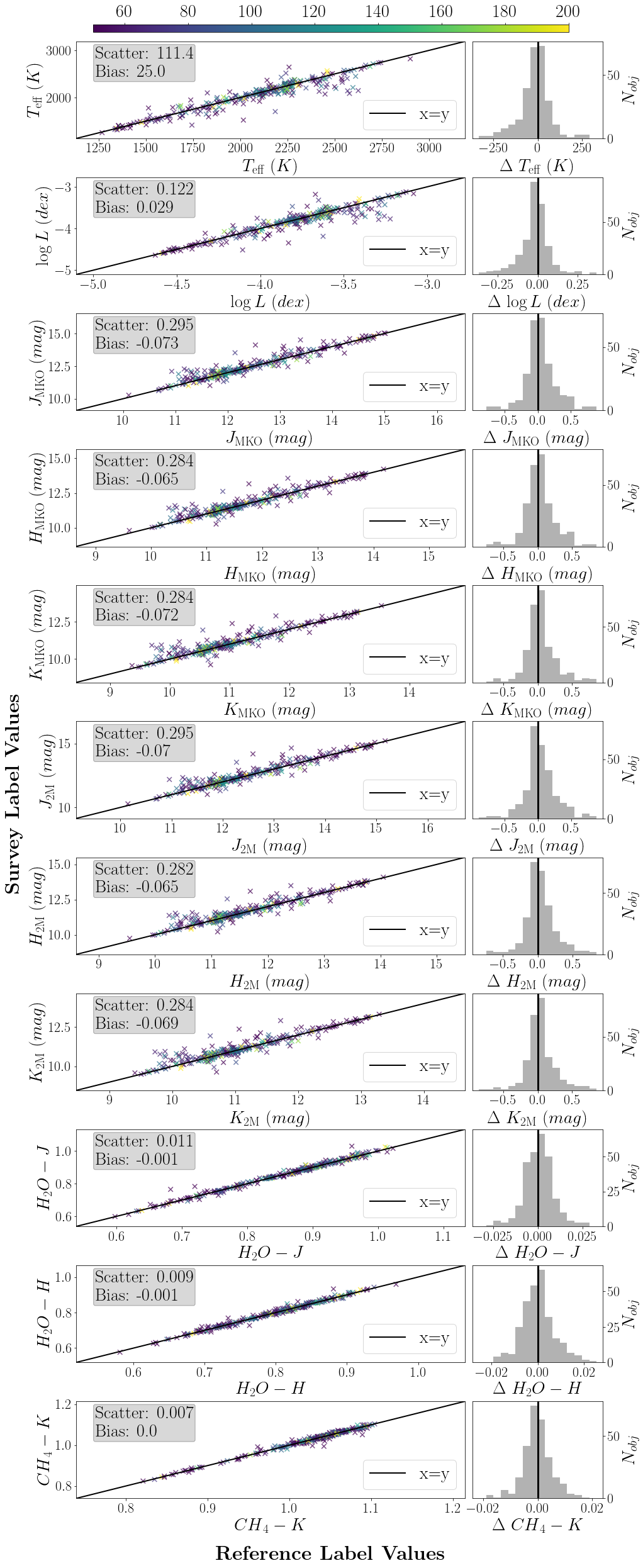}}\quad
    \subfloat[{\TC} Model $M_{4T}$ for L8--T8.5 dwarfs]{\includegraphics[scale=0.245]{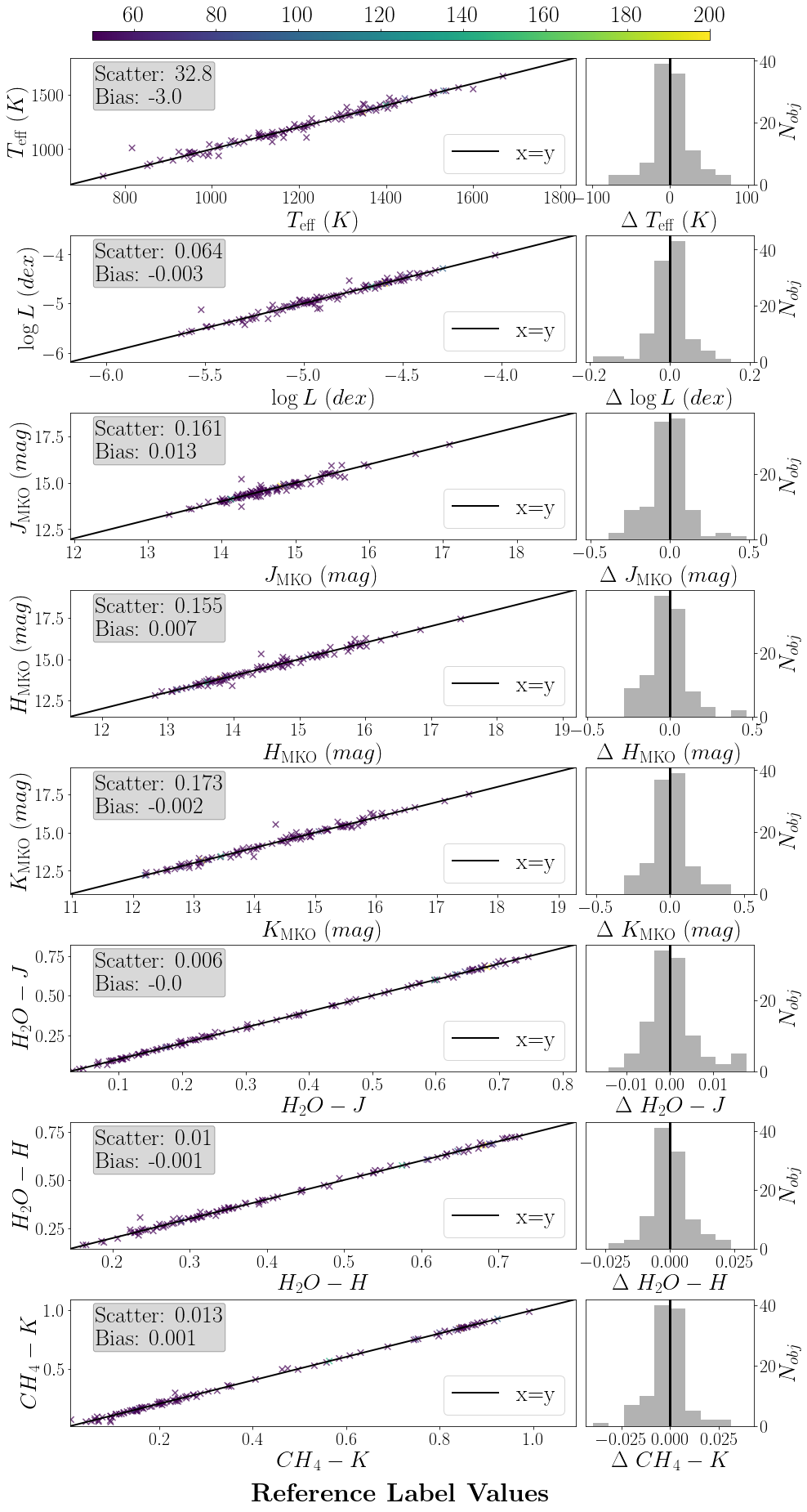}}
    \caption{Same as Figure~\ref{fig:Lab3b} but comparing the reference and survey values for {\TC} models $M_{4L}$ and $M_{4T}$ (Section~\ref{sec:fin_mod}) that use distinct reference sets containing L0--L8 (left) and L8--T8.5 (right). This produces the lowest scatter and bias for all labels, demonstrating that {\TC} infers more accurate physical quantities when limiting the spectral type ranges of the reference sets used to generate the model.}
    \label{fig:Lab4}
\end{figure*}

Figures~\ref{fig:Lab4} compares the inferred and reference values for these models. The RMS of the inferred labels for the L dwarf model is nominally reduced, but for T dwarfs the reduction in RMS is significant, with a 53~per~cent decrease for {\Teff}, 24~per~cent decrease for {\logL}, and 43~per~cent decrease for $JHK_{\rm MKO}$ photometry. This indicates that separating L and T dwarfs in the reference set and utilizing spectral indices rather than spectral types produces the most accurate models with {\TC}. This also shows that {\TC} can effectively transfer labels from reference sets as small as $\approx$100 objects, and that the modest S/N of the T~dwarf spectra (typically 20--50) is not an impediment. Notably, the $M_{4L}$ model produces labels for L~dwarfs with higher RMS and bias, even though the L~dwarf reference set is larger and the S/N are typically higher than in the T~dwarf reference set.

\subsubsection{Discussion of Individual Objects of Interest}
During our construction of {\TC} models described in this Section, we identified outlier objects, which we define as having inferred label values more than two standard deviations from the reference label space in more than one model. 
We identified 10 such objects, most of which are L0~dwarfs with no unusual spectral features or other properties noted in the literature. 
We suspect that {\TC} is returning less accurate survey labels for these objects because the lie near the edges of our reference label spaces.
Three objects do have unusual properties which we discuss here.

\textit{2MASSI J0859254$-$194926}: This object is categorized as an optical L6~dwarf by \citet[adopted by us]{Cruz:2003fi} and a NIR L8$\pm2$~dwarf by \citet{Thompson:2013kv}. It is a faint outlier in absolute magnitude for an L6~dwarf in 2MASS and WISE bands (Figure~\ref{fig:poly_2mass-wise}). {\TC} models that include {\SpT} as a label return inferred values that are 1.2--2 spectral types greater than the reference value. We conclude that this object is best classified as an L8 dwarf.

\textit{DENIS J1707252$-$013809}: This L2~dwarf \citep{Martin:2010cx} is a weak binary candidate whose spectrum best fits components with L0.7 and T4.3 spectral types \citep{BardalezGagliuffi2014}. {\TC} models that use both L and T dwarfs in the reference set infer values for {\Teff}, {\logL}, and photometry labels more than $2\sigma$ cooler and fainter than the reference label space, and {\TC} models that include {\SpT} as a label return inferred values that are 1.5--2 spectral subtypes cooler than the reference value. However, when the reference set is restricted to L dwarfs, the inferred labels are consistent with reference labels. Including T dwarfs in the reference set could allow {\TC} to identify spectroscopic features that correspond to lower temperatures and fainter magnitudes.

\textit{SDSS J104409.43+042937.6}: This object is $\approx$0.4~mag overluminous for its L7 spectral type \citep{Knapp2004a} in the near-infrared photometric bands when compared to its parallax-derived distance. It has no unusual spectroscopic features, so we consider it to be a candidate $\approx$L7+L7 binary. {\TC} models infer values for {\Teff}, {\logL}, and photometry labels more than $2\sigma$ cooler and fainter than the reference label space for both the $M_3$ model (L and T dwarfs in the reference set) and $M_{4L}$ model (only L~dwarfs in the reference set).

\begin{figure*}
    \centering
    \subfloat[{\TC} Model $M_{4L, R\sim40}$ for L0--L8 dwarfs]{\includegraphics[scale=0.245]{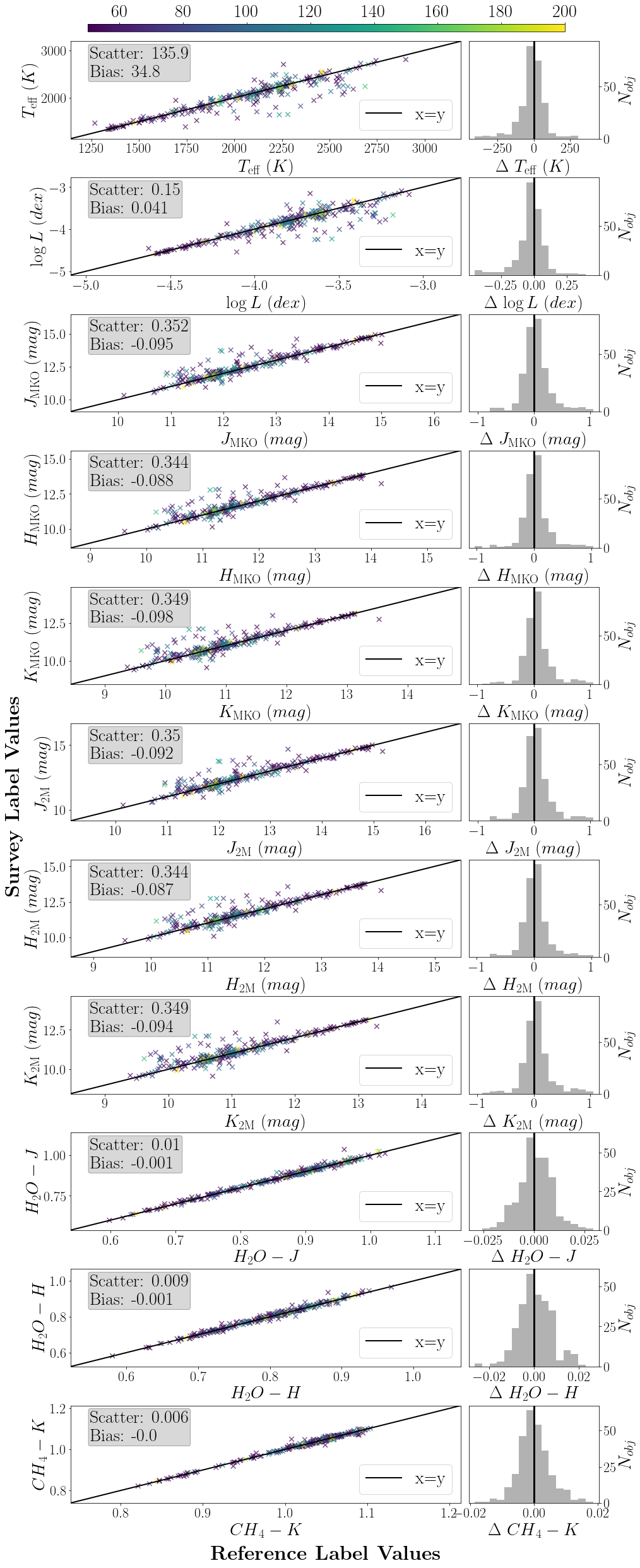}}\quad
    \subfloat[{\TC} $M_{4T, R\sim40}$ for L8--T8.5 dwarfs]{\includegraphics[scale=0.245]{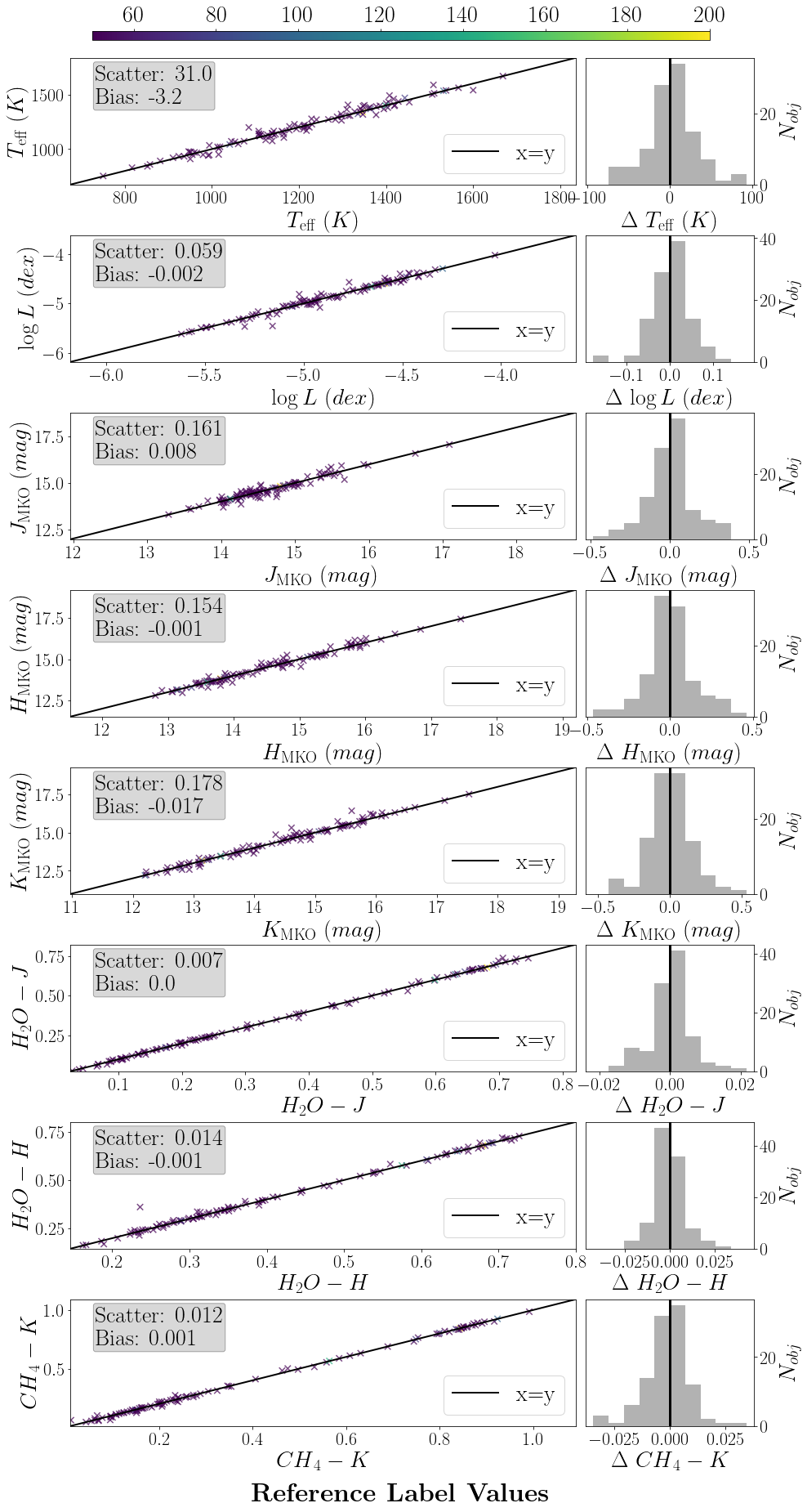}}
    \caption{Same as Figure~\ref{fig:Lab4} but comparing the reference and survey values for {\TC} models that utilize spectra at a resolution $R\sim40$ (Section~\ref{sec:futureapp}). These models use the same reference set and labels of {\TC} models $M_{4L}$ (L0--L8, left) and $M_{4T}$ (L8--T8.5, right). Using lower-resolution spectra, {\TC} can determine labels for L dwarfs with $\approx$20\% less precision but infers labels for T dwarfs at the same or better precision.}
    \label{fig:Lab5}
\end{figure*}

\subsection{Application to SPHEREx-resolution Spectra}
\label{sec:futureapp}
We briefly explored the efficacy of {\TC} to infer the same physical properties from spectra taken by future space missions. The Spectro-Photometer for the History of the Universe, Epoch of Reionization and Ices Explorer \citep[SPHEREx;][]{Crill:2020ep} spacecraft, currently scheduled for launch in 2025, will survey the sky at wavelengths $0.75<\lambda<5.0~\mu$m. For near-IR wavelengths ($0.75<\lambda<2.4~\mu$m), the spectral resolution will be $R=41$ \citep{Dore:2014}, about half that of the SpeX prism spectra we have analysed in this work. To simulate SPHEREx spectra, we removed every other element (i.e., 220 of the 440 elements) from the wavelength grid of the same $R\sim75$ spectrum we used in Section~\ref{sec:prep}, re-interpolated the spectra onto this grid to obtain $R\sim40$, and reevaluated {\TC} models described in Section~\ref{sec:TC_res}.

For each model, we used the same labels, reference sets, and applied the same continuum estimation (i.e., wavelength ranges and percentile cutoffs) as described previously. The identified continuum pixels were fit using a first-order sinusoid function. When comparing the $R\sim40$ spectra models to their respective $R\sim75$ spectra models ({\TC} $M_1$, $M_2$, $M_3$) all models that used the lower resolution spectra in the reference set resulted in an increase in scatter for {\Teff} (5 to 8 per cent), {\logL} (4 to 8 per cent), and ${MKO}$ and ${2MASS}$ photometry (2 to 9 per cent). Figure~\ref{fig:Lab5} shows the results that use lower resolution spectra for {\TC} models $M_{4L}$ and $M_{4T}$ (Section~\ref{sec:fin_mod}), which separate L and T dwarfs in the reference sets. Using lower resolution spectra in the $M_{4L}$ model, L0--L8 dwarfs reference set, increases the scatter 18 to 23 per cent for {\Teff}, {\logL}, and all ${MKO}$ and ${2MASS}$ bands. For the $M_{4T}$ model, L8--T8.5 dwarfs reference set, we note a 5 to 8 per cent decrease in the scatter for {\Teff} and {\logL}, no change for $J_{MKO}$ and $H_{MKO}$ photometry labels, and a 3 per cent increase in scatter for the $K_{MKO}$ label. {\TC} can determine labels for L dwarfs with lower precision, but infers physical properties for T dwarfs at the same or slightly greater precision. This indicates that physical properties can be inferred by {\TC} from a reference set of lower-resolution $R\sim40$ spectra and, providing accurate reference labels, {\TC} should be useful for $R\sim40$ spectra that are expected to come from SPHEREx.

\section{Discussion}
We have demonstrated that the machine learning tool {\TC} can be used to accurately infer physical quantities (``labels'') such as effective temperature, photometry, and bolometric luminosity from low-resolution ({$R$}$\sim$75) SpeX prism spectra of L and T dwarfs. Previous applications of {\TC} have analysed high-resolution spectra of stars that could be consistently normalized to a continuum. We have proven that physical information can similarly be extracted from the spectra of brown dwarfs, which are characterized by broad molecular absorption features and a lack of continuum.  

{\TC} can infer numerical spectral types, absolute magnitudes, effective temperatures, and bolometric luminosities at precisions that are competitive with other methods, without the need for new parallax measurements. We find that using spectral indices instead of numerical spectral types for reference labels improves (i.e., decreases) the scatter and bias of the inferred values for all survey labels. Our best models can determine labels for L~dwarfs with precisions of $\approx$100~K in {\Teff}, $\approx$0.12~dex in {\logL}, and $\lesssim$0.30~mag in MKO and 2MASS $JHK$ photometry. Similarly, properties of T dwarfs can be inferred with precisions of $\approx$30~K in {\Teff}, $\approx$0.06~dex in {\logL}, and $\lesssim$0.17~mag in MKO $JHK$ photometry. Spectral indices can also be determined at precisions of $\approx0.01$. These precisions for {\Teff} and {\logL} in T~dwarfs are comparable to methods that employ absolute magnitude-based polynomial relations \citep[e.g.,][]{Filippazzo2015,Dupuy2017}, indicating that the {\TC} can reach the systematic uncertainty limit in the polynomial-determined reference labels.
When comparing inferred labels to the absolute magnitudes determined with spectral type-based relations, we find {\TC} can infer photometry at lower uncertainties and without requiring trigonometric parallaxes. We briefly examine the efficacy of {\TC} models for future spectroscopic missions. Replicating the resolving power of SPHEREx, $R\sim40$, we predict {\TC} will successfully infer physical properties with similar or lower precision than models that use higher-resolution spectra. We have not yet fully explored the possible combinations of data sets and parameters for {\TC}, and we expect future work to improve upon our results.

We have also investigated several factors that affect {\TC}'s ability to infer accurate labels. First, {\TC} relies on a continuum estimation that is used to identify wavelength regions in the spectra that do not change significantly with the labels and are near unity flux (a ``continuum mask''). Typically, regions that are strongly affected by absorption are not identified for {\TC}'s continuum normalization. However, applying this continuum mask for brown dwarf spectra is challenging since they contain few or no absorption-free regions. We found that typically, identifying $J$, $H$, and $K$ bands as three distinct wavelength regions to be continuum-normalized separately is most appropriate for this process. Second, the accuracy of the inferred label values are limited by the quality of the reference labels given to train the model. We have assumed that the reference labels used for these models are accurate, but in reality they contain some uncertainty and potentially some bias. Including label uncertainties in {\TC} models as well as utilizing tailored methods to obtain more accurate reference label values (e.g., {\Teff} and {\logL} from spectral energy distribution fitting) could improve the inference of properties for new spectra. In future work, label uncertainties could be used as variables in the label functions or as labels themselves.

\section*{Acknowledgements}
We thank Anna Ho for useful discussions regarding {\TC} and Michael Gully-Santiago for helpful conversations about machine learning algorithms. 
This work has benefitted from The UltracoolSheet at \url{http://bit.ly/UltracoolSheet}, maintained by Will Best, Trent Dupuy, Michael Liu, Rob Siverd, and Zhoujian Zhang, and developed from compilations by \citet{Dupuy2012}, \citet{Dupuy:2013ks}, \citet{LiuDupuy2016}, \citet{Best2018}, and \citet{Best2021}.		
This research has benefitted from the SpeX Prism Library and the SpeX Prism Library Analysis Toolkit, maintained by Adam Burgasser at \url{http://www.browndwarfs.org/spexprism}.
This work has made use of data from the European Space Agency (ESA) mission Gaia (\url{http://www.cosmos.esa.int/gaia}), processed by the Gaia Data Processing and Analysis Consortium (DPAC, \url{http://www.cosmos.esa.int/web/gaia/dpac/consortium}). Funding for the DPAC has been provided by national institutions, in particular the institutions participating in the Gaia Multilateral Agreement.
This publication makes use of data products from the Two Micron All Sky Survey (2MASS), which is a joint project of the University of Massachusetts and the Infrared Processing and Analysis Center/California Institute of Technology, funded by the National Aeronautics and Space Administration and the National Science Foundation.
The Pan-STARRS1 Surveys (PS1) and the PS1 public science archive have been made possible through contributions by the Institute for Astronomy, the University of Hawaii, the Pan-STARRS Project Office, the Max-Planck Society and its participating institutes, the Max Planck Institute for Astronomy, Heidelberg and the Max Planck Institute for Extraterrestrial Physics, Garching, The Johns Hopkins University, Durham University, the University of Edinburgh, the Queen's University Belfast, the Harvard-Smithsonian Center for Astrophysics, the Las Cumbres Observatory Global Telescope Network Incorporated, the National Central University of Taiwan, the Space Telescope Science Institute, the National Aeronautics and Space Administration under grant No. NNX08AR22G issued through the Planetary Science Division of the NASA Science Mission Directorate, the National Science Foundation grant No. AST-1238877, the University of Maryland, Eotvos Lorand University (ELTE), the Los Alamos National Laboratory, and the Gordon and Betty Moore Foundation.
This publication makes use of data products from the Wide-field Infrared Survey Explorer, which is a joint project of the University of California, Los Angeles, and the Jet Propulsion Laboratory/California Institute of Technology, and NEOWISE, which is a project of the Jet Propulsion Laboratory/California Institute of Technology. WISE and NEOWISE are funded by the National Aeronautics and Space Administration. This research has made use of NASA's Astrophysical Data System and the SIMBAD and Vizier databases operated at CDS, Strasbourg, France.
S.J.F. acknowledges the generous support from the John W. Cox Endowment for the Advanced Studies in Astronomy and the TIDES Advanced Research Fellowship.
W.M.J.B. received support from grant HST-GO-15238 provided by STScI and AURA.
Finally, the authors wish to recognize and acknowledge the very significant cultural role and reverence that the summit of Maunakea has always held within the indigenous Hawaiian community. We are most fortunate to have the opportunity to work with data from this mountain.

\section*{Data Availability}
The data underlying the absolute magnitude vs. spectral type polynomial relations derived in Section~\ref{sec:poly} are available in the UltracoolSheet at \url{https://bit.ly/UltracoolSheet/} and in Zenodo at \url{https://doi.org/10.5281/zenodo.4169084}.
The data underlying {\TC} models are available in the SpeX Prism Library at \url{http://pono.ucsd.edu/~adam/browndwarfs/spexprism/}.
New data products produced for this article will be shared on reasonable request to the corresponding author.

\bibliographystyle{mnras}
\bibliography{BD_TC}

\begin{thebibliography}{}
\makeatletter
\relax
\def\mn@urlcharsother{\let\do\@makeother \do\$\do\&\do\#\do\^\do\_\do\%\do\~}
\def\mn@doi{\begingroup\mn@urlcharsother \@ifnextchar [ {\mn@doi@}
  {\mn@doi@[]}}
\def\mn@doi@[#1]#2{\def\@tempa{#1}\ifx\@tempa\@empty \href
  {http://dx.doi.org/#2} {doi:#2}\else \href {http://dx.doi.org/#2} {#1}\fi
  \endgroup}
\def\mn@eprint#1#2{\mn@eprint@#1:#2::\@nil}
\def\mn@eprint@arXiv#1{\href {http://arxiv.org/abs/#1} {{\tt arXiv:#1}}}
\def\mn@eprint@dblp#1{\href {http://dblp.uni-trier.de/rec/bibtex/#1.xml}
  {dblp:#1}}
\def\mn@eprint@#1:#2:#3:#4\@nil{\def\@tempa {#1}\def\@tempb {#2}\def\@tempc
  {#3}\ifx \@tempc \@empty \let \@tempc \@tempb \let \@tempb \@tempa \fi \ifx
  \@tempb \@empty \def\@tempb {arXiv}\fi \@ifundefined
  {mn@eprint@\@tempb}{\@tempb:\@tempc}{\expandafter \expandafter \csname
  mn@eprint@\@tempb\endcsname \expandafter{\@tempc}}}

\bibitem[\protect\citeauthoryear{Ackerman \& Marley}{Ackerman \&
  Marley}{2001}]{Ackerman2001}
Ackerman A.~S.,  Marley M.~S.,  2001, \mn@doi [ApJ] {10.1086/321540}, 556, 872

\bibitem[\protect\citeauthoryear{{Aganze} et~al.,}{{Aganze}
  et~al.}{2021}]{Aganze:2021}
{Aganze} C.,  et~al., 2021, arXiv e-prints, p. arXiv:2110.07672

\bibitem[\protect\citeauthoryear{Albert, Artigau, Delorme, Reyl{\'e},
  Forveille, Delfosse  \& Willott}{Albert et~al.}{2011}]{Albert:2011bc}
Albert L.,  Artigau E.,  Delorme P.,  Reyl{\'e} C.,  Forveille T.,  Delfosse
  X.,   Willott C.~J.,  2011, AJ, 141, 203

\bibitem[\protect\citeauthoryear{Allers \& Liu}{Allers \&
  Liu}{2013}]{Allers2013}
Allers K.~N.,  Liu M.~C.,  2013, \mn@doi [ApJ] {10.1088/0004-637X/772/2/79},
  772, 79

\bibitem[\protect\citeauthoryear{Allers et~al.,}{Allers
  et~al.}{2007}]{Allers2007}
Allers K.~N.,  et~al., 2007, \mn@doi [ApJ] {10.1086/510845}, 657, 511

\bibitem[\protect\citeauthoryear{{Bardalez Gagliuffi} et~al.,}{{Bardalez
  Gagliuffi} et~al.}{2014}]{BardalezGagliuffi2014}
{Bardalez Gagliuffi} D.~C.,  et~al., 2014, \mn@doi [ApJ]
  {10.1088/0004-637X/794/2/143}, 794, 143

\bibitem[\protect\citeauthoryear{Behmard, Petigura  \& Howard}{Behmard
  et~al.}{2019}]{Behmard2019}
Behmard A.,  Petigura E.~A.,   Howard A.~W.,  2019, \mn@doi [ApJ]
  {10.3847/1538-4357/ab14e0}, 876, 68

\bibitem[\protect\citeauthoryear{Beichman, Gelino, Kirkpatrick, Cushing,
  Dodson-Robinson, Marley, Morley  \& Wright}{Beichman
  et~al.}{2014}]{Beichman:2014jr}
Beichman C.,  Gelino C.~R.,  Kirkpatrick J.~D.,  Cushing M.~C.,
  Dodson-Robinson S.,  Marley M.~S.,  Morley C.~V.,   Wright E.~L.,  2014, ApJ,
  783, 68

\bibitem[\protect\citeauthoryear{Best et~al.,}{Best et~al.}{2013}]{Best:2013bp}
Best W. M.~J.,  et~al., 2013, ApJ, 777, 84

\bibitem[\protect\citeauthoryear{Best et~al.,}{Best et~al.}{2015}]{Best:2015em}
Best W. M.~J.,  et~al., 2015, ApJ, 814, 118

\bibitem[\protect\citeauthoryear{Best et~al.,}{Best et~al.}{2018}]{Best2018}
Best W. M.~J.,  et~al., 2018, \mn@doi [ApJS] {10.3847/1538-4365/aa9982}, 234, 1

\bibitem[\protect\citeauthoryear{Best, Liu, Magnier  \& Dupuy}{Best
  et~al.}{2020}]{Best2020a}
Best W. M.~J.,  Liu M.~C.,  Magnier E.~A.,   Dupuy T.~J.,  2020, \mn@doi [AJ]
  {10.3847/1538-3881/ab84f4}, 159, 257

\bibitem[\protect\citeauthoryear{Best, Liu, Magnier  \& Dupuy}{Best
  et~al.}{2021}]{Best2021}
Best W. M.~J.,  Liu M.~C.,  Magnier E.~A.,   Dupuy T.~J.,  2021, \mn@doi [AJ]
  {10.3847/1538-3881/abc893}, 161, 42

\bibitem[\protect\citeauthoryear{Bihain, Scholz, Storm  \& Schnurr}{Bihain
  et~al.}{2013}]{Bihain:2013gw}
Bihain G.,  Scholz R.-D.,  Storm J.,   Schnurr O.,  2013, A{\&}A, 557, 43

\bibitem[\protect\citeauthoryear{Boccaletti, Chauvin, Lagrange  \&
  Marchis}{Boccaletti et~al.}{2003}]{Boccaletti:2003cl}
Boccaletti A.,  Chauvin G.,  Lagrange A.~M.,   Marchis F.,  2003, A{\&}A, 410,
  283

\bibitem[\protect\citeauthoryear{Burgasser}{Burgasser}{2014}]{Burgasser2014}
Burgasser A.~J.,  2014, in Singh H.~P.,  Prugniel P.,   Vauglin I.,  eds, ASI
  Conference Series 11. Bangalore: Astronomical Society of India, pp 7--16

\bibitem[\protect\citeauthoryear{Burgasser et~al.,}{Burgasser
  et~al.}{1999}]{Burgasser:1999fp}
Burgasser A.~J.,  et~al., 1999, ApJ, 522, L65

\bibitem[\protect\citeauthoryear{Burgasser et~al.,}{Burgasser
  et~al.}{2000}]{Burgasser:2000bm}
Burgasser A.~J.,  et~al., 2000, ApJ, 531, L57

\bibitem[\protect\citeauthoryear{Burgasser et~al.,}{Burgasser
  et~al.}{2002a}]{Burgasser:2002fy}
Burgasser A.~J.,  et~al., 2002a, ApJ, 564, 421

\bibitem[\protect\citeauthoryear{Burgasser, Marley, Ackerman, Saumon, Lodders,
  Dahn, Harris  \& Kirkpatrick}{Burgasser et~al.}{2002b}]{Burgasser2002}
Burgasser A.~J.,  Marley M.~S.,  Ackerman A.~S.,  Saumon D.,  Lodders K.,  Dahn
  C.~C.,  Harris H.~C.,   Kirkpatrick J.~D.,  2002b, \mn@doi [ApJ]
  {10.1086/341343}, 571, L151

\bibitem[\protect\citeauthoryear{Burgasser, Kirkpatrick, McElwain, Cutri,
  Burgasser  \& Skrutskie}{Burgasser et~al.}{2003a}]{Burgasser:2003ij}
Burgasser A.~J.,  Kirkpatrick J.~D.,  McElwain M.~W.,  Cutri R.~M.,  Burgasser
  A.~J.,   Skrutskie M.~F.,  2003a, AJ, 125, 850

\bibitem[\protect\citeauthoryear{Burgasser, McElwain  \& Kirkpatrick}{Burgasser
  et~al.}{2003b}]{Burgasser:2003jf}
Burgasser A.~J.,  McElwain M.~W.,   Kirkpatrick J.~D.,  2003b, AJ, 126, 2487

\bibitem[\protect\citeauthoryear{Burgasser, Kirkpatrick, Liebert  \&
  Burrows}{Burgasser et~al.}{2003c}]{Burgasser:2003dh}
Burgasser A.~J.,  Kirkpatrick J.~D.,  Liebert J.,   Burrows A.~S.,  2003c, ApJ,
  594, 510

\bibitem[\protect\citeauthoryear{Burgasser, McElwain, Kirkpatrick, Cruz, Tinney
   \& Reid}{Burgasser et~al.}{2004}]{Burgasser:2004hg}
Burgasser A.~J.,  McElwain M.~W.,  Kirkpatrick J.~D.,  Cruz K.~L.,  Tinney
  C.~G.,   Reid I.~N.,  2004, AJ, 127, 2856

\bibitem[\protect\citeauthoryear{Burgasser, Geballe, Leggett, Kirkpatrick  \&
  Golimowski}{Burgasser et~al.}{2006}]{Burgasser2006}
Burgasser A.~J.,  Geballe T.~R.,  Leggett S.~K.,  Kirkpatrick J.~D.,
  Golimowski D.~A.,  2006, \mn@doi [ApJ] {10.1086/498563}, 637, 1067

\bibitem[\protect\citeauthoryear{Burgasser, Looper, Kirkpatrick, Cruz  \&
  Swift}{Burgasser et~al.}{2008a}]{Burgasser:2008ei}
Burgasser A.~J.,  Looper D.~L.,  Kirkpatrick J.~D.,  Cruz K.~L.,   Swift B.~J.,
   2008a, ApJ, 674, 451

\bibitem[\protect\citeauthoryear{Burgasser, Liu, Ireland, Cruz  \&
  Dupuy}{Burgasser et~al.}{2008b}]{Burgasser:2008cj}
Burgasser A.~J.,  Liu M.~C.,  Ireland M.~J.,  Cruz K.~L.,   Dupuy T.~J.,
  2008b, ApJ, 681, 579

\bibitem[\protect\citeauthoryear{Burgasser, Tinney, Cushing, Saumon, Marley,
  Bennett  \& Kirkpatrick}{Burgasser et~al.}{2008c}]{Burgasser:2008ke}
Burgasser A.~J.,  Tinney C.~G.,  Cushing M.~C.,  Saumon D.,  Marley M.~S.,
  Bennett C.~S.,   Kirkpatrick J.~D.,  2008c, ApJL, 689, L53

\bibitem[\protect\citeauthoryear{Burgasser, Cruz, Cushing, Gelino, Looper,
  Faherty, Kirkpatrick  \& Reid}{Burgasser et~al.}{2010}]{Burgasser2010}
Burgasser A.~J.,  Cruz K.~L.,  Cushing M.,  Gelino C.~R.,  Looper D.~L.,
  Faherty J.~K.,  Kirkpatrick J.~D.,   Reid I.~N.,  2010, \mn@doi [ApJ]
  {10.1088/0004-637X/710/2/1142}, 710, 1142

\bibitem[\protect\citeauthoryear{Burningham et~al.,}{Burningham
  et~al.}{2008}]{Burningham:2008fc}
Burningham B.,  et~al., 2008, MNRAS, 391, 320

\bibitem[\protect\citeauthoryear{Burningham et~al.,}{Burningham
  et~al.}{2009}]{Burningham:2009ft}
Burningham B.,  et~al., 2009, MNRAS, 395, 1237

\bibitem[\protect\citeauthoryear{Burningham et~al.,}{Burningham
  et~al.}{2010}]{Burningham:2010dh}
Burningham B.,  et~al., 2010, MNRAS, 406, 1885

\bibitem[\protect\citeauthoryear{Burningham et~al.,}{Burningham
  et~al.}{2011}]{Burningham:2011kh}
Burningham B.,  et~al., 2011, MNRASL, 414, L90

\bibitem[\protect\citeauthoryear{Burningham et~al.,}{Burningham
  et~al.}{2013}]{Burningham:2013gt}
Burningham B.,  et~al., 2013, MNRAS, 433, 457

\bibitem[\protect\citeauthoryear{Burrows et~al.,}{Burrows
  et~al.}{1997}]{Burrows1997}
Burrows A.,  et~al., 1997, \mn@doi [ApJ] {10.1086/305002}, 491, 856

\bibitem[\protect\citeauthoryear{Casey et~al.,}{Casey et~al.}{2017}]{Casey2017}
Casey A.~R.,  et~al., 2017, \mn@doi [ApJ] {10.3847/1538-4357/aa69c2}, 840, 59

\bibitem[\protect\citeauthoryear{Castro \& Gizis}{Castro \&
  Gizis}{2012}]{Castro:2012dj}
Castro P.~J.,  Gizis J.~E.,  2012, ApJ, 746, 3

\bibitem[\protect\citeauthoryear{Castro, Gizis, Harris, Mace, Kirkpatrick,
  McLean, Pattarakijwanich  \& Skrutskie}{Castro et~al.}{2013}]{Castro:2013bb}
Castro P.~J.,  Gizis J.~E.,  Harris H.~C.,  Mace G.~N.,  Kirkpatrick J.~D.,
  McLean I.~S.,  Pattarakijwanich P.,   Skrutskie M.~F.,  2013, ApJ, 776, 126

\bibitem[\protect\citeauthoryear{Chambers et~al.,}{Chambers
  et~al.}{2021}]{Chambers2016}
Chambers K.~C.,  et~al., 2021, ApJS, in press, arXiv:1612.05560

\bibitem[\protect\citeauthoryear{Chiu, Fan, Leggett, Golimowski, Zheng,
  Geballe, Schneider  \& Brinkmann}{Chiu et~al.}{2006}]{Chiu:2006jd}
Chiu K.,  Fan X.,  Leggett S.~K.,  Golimowski D.~A.,  Zheng W.,  Geballe T.~R.,
   Schneider D.~P.,   Brinkmann J.,  2006, AJ, 131, 2722

\bibitem[\protect\citeauthoryear{Crill et~al.,}{Crill
  et~al.}{2020}]{Crill:2020ep}
Crill B.~P.,  et~al., 2020, in Proc. SPIE. p. 114430I

\bibitem[\protect\citeauthoryear{Cruz, Reid, Liebert, Kirkpatrick  \&
  Lowrance}{Cruz et~al.}{2003}]{Cruz:2003fi}
Cruz K.~L.,  Reid I.~N.,  Liebert J.,  Kirkpatrick J.~D.,   Lowrance P.~J.,
  2003, AJ, 126, 2421

\bibitem[\protect\citeauthoryear{Cruz et~al.,}{Cruz et~al.}{2007}]{Cruz:2007kb}
Cruz K.~L.,  et~al., 2007, AJ, 133, 439

\bibitem[\protect\citeauthoryear{Cushing et~al.,}{Cushing
  et~al.}{2011}]{Cushing:2011dk}
Cushing M.~C.,  et~al., 2011, ApJ, 743, 50

\bibitem[\protect\citeauthoryear{Cushing, Kirkpatrick, Gelino, Mace, Skrutskie
  \& Gould}{Cushing et~al.}{2014}]{Cushing:2014be}
Cushing M.~C.,  Kirkpatrick J.~D.,  Gelino C.~R.,  Mace G.~N.,  Skrutskie
  M.~F.,   Gould A.,  2014, AJ, 147, 113

\bibitem[\protect\citeauthoryear{Cutri et~al.,}{Cutri
  et~al.}{2003}]{Cutri:2003vr}
Cutri R.~M.,  et~al., 2003, yCat, II/246, 0

\bibitem[\protect\citeauthoryear{Cutri et~al.,}{Cutri
  et~al.}{2014}]{Cutri:2014wx}
Cutri R.~M.,  et~al., 2014, yCat, II/328, 0

\bibitem[\protect\citeauthoryear{Dahn et~al.,}{Dahn et~al.}{2017}]{Dahn:2017gu}
Dahn C.~C.,  et~al., 2017, AJ, 154, 147

\bibitem[\protect\citeauthoryear{Deacon et~al.,}{Deacon
  et~al.}{2011}]{Deacon:2011gz}
Deacon N.~R.,  et~al., 2011, AJ, 142, 77

\bibitem[\protect\citeauthoryear{Deacon et~al.,}{Deacon
  et~al.}{2012a}]{Deacon:2012eg}
Deacon N.~R.,  et~al., 2012a, ApJ, 755, 94

\bibitem[\protect\citeauthoryear{Deacon et~al.,}{Deacon
  et~al.}{2012b}]{Deacon:2012gf}
Deacon N.~R.,  et~al., 2012b, ApJ, 757, 100

\bibitem[\protect\citeauthoryear{Deacon et~al.,}{Deacon
  et~al.}{2014}]{Deacon:2014ey}
Deacon N.~R.,  et~al., 2014, ApJ, 792, 119

\bibitem[\protect\citeauthoryear{Deacon et~al.,}{Deacon
  et~al.}{2017}]{Deacon:2017kd}
Deacon N.~R.,  et~al., 2017, MNRAS, 467, 1126

\bibitem[\protect\citeauthoryear{Delfosse et~al.,}{Delfosse
  et~al.}{1997}]{Delfosse:1997uj}
Delfosse X.,  et~al., 1997, A{\&}A, 327, L25

\bibitem[\protect\citeauthoryear{Delfosse, Tinney, Forveille, Epchtein,
  Borsenberger, Fouque, Kimeswenger  \& Tiph{\`e}ne}{Delfosse
  et~al.}{1999}]{Delfosse:1999bx}
Delfosse X.,  Tinney C.~G.,  Forveille T.,  Epchtein N.,  Borsenberger J.,
  Fouque P.,  Kimeswenger S.,   Tiph{\`e}ne D.,  1999, A{\&}AS, 135, 41

\bibitem[\protect\citeauthoryear{Delorme et~al.,}{Delorme
  et~al.}{2008}]{Delorme:2008jd}
Delorme P.,  et~al., 2008, A{\&}A, 482, 961

\bibitem[\protect\citeauthoryear{{Dor{\'e}} et~al.,}{{Dor{\'e}}
  et~al.}{2014}]{Dore:2014}
{Dor{\'e}} O.,  et~al., 2014, arXiv e-prints, p. arXiv:1412.4872

\bibitem[\protect\citeauthoryear{Dupuy \& Kraus}{Dupuy \&
  Kraus}{2013}]{Dupuy:2013ks}
Dupuy T.~J.,  Kraus A.~L.,  2013, Science, 341, 1492

\bibitem[\protect\citeauthoryear{Dupuy \& Liu}{Dupuy \& Liu}{2012}]{Dupuy2012}
Dupuy T.~J.,  Liu M.~C.,  2012, \mn@doi [ApJS] {10.1088/0067-0049/201/2/19},
  201, 19

\bibitem[\protect\citeauthoryear{Dupuy \& Liu}{Dupuy \& Liu}{2017}]{Dupuy2017}
Dupuy T.~J.,  Liu M.~C.,  2017, \mn@doi [ApJS] {10.3847/1538-4365/aa5e4c}, 231,
  15

\bibitem[\protect\citeauthoryear{Faherty et~al.,}{Faherty
  et~al.}{2012}]{Faherty:2012cy}
Faherty J.~K.,  et~al., 2012, ApJ, 752, 56

\bibitem[\protect\citeauthoryear{Faherty et~al.,}{Faherty
  et~al.}{2016}]{Faherty2016a}
Faherty J.~K.,  et~al., 2016, \mn@doi [ApJS] {10.3847/0067-0049/225/1/10}, 225,
  10

\bibitem[\protect\citeauthoryear{Fan et~al.,}{Fan et~al.}{2000}]{Fan:2000iu}
Fan X.,  et~al., 2000, AJ, 119, 928

\bibitem[\protect\citeauthoryear{Filippazzo, Rice, Faherty, Cruz, Gordon  \&
  Looper}{Filippazzo et~al.}{2015}]{Filippazzo2015}
Filippazzo J.~C.,  Rice E.~L.,  Faherty J.,  Cruz K.~L.,  Gordon M. M.~V.,
  Looper D.~L.,  2015, \mn@doi [ApJ] {10.1088/0004-637X/810/2/158}, 810, 158

\bibitem[\protect\citeauthoryear{Gagn{\'e} et~al.,}{Gagn{\'e}
  et~al.}{2015}]{Gagne:2015dc}
Gagn{\'e} J.,  et~al., 2015, ApJS, 219, 33

\bibitem[\protect\citeauthoryear{{Gaia Collaboration} et~al.,}{{Gaia
  Collaboration} et~al.}{2018}]{GaiaCollaboration:2018io}
{Gaia Collaboration} et~al., 2018, A{\&}A, 616, A1

\bibitem[\protect\citeauthoryear{Geballe et~al.,}{Geballe
  et~al.}{2002}]{Geballe:2002kw}
Geballe T.~R.,  et~al., 2002, ApJ, 564, 466

\bibitem[\protect\citeauthoryear{Gizis}{Gizis}{2002}]{Gizis:2002je}
Gizis J.~E.,  2002, ApJ, 575, 484

\bibitem[\protect\citeauthoryear{Gizis, Monet, Reid, Kirkpatrick, Liebert  \&
  Williams}{Gizis et~al.}{2000}]{Gizis:2000kz}
Gizis J.~E.,  Monet D.~G.,  Reid I.~N.,  Kirkpatrick J.~D.,  Liebert J.,
  Williams R.~J.,  2000, AJ, 120, 1085

\bibitem[\protect\citeauthoryear{Gizis, Kirkpatrick  \& Wilson}{Gizis
  et~al.}{2001}]{Gizis:2001jp}
Gizis J.~E.,  Kirkpatrick J.~D.,   Wilson J.~C.,  2001, AJ, 121, 2185

\bibitem[\protect\citeauthoryear{Gizis, Burgasser, Faherty, Castro  \&
  Shara}{Gizis et~al.}{2011a}]{Gizis:2011jv}
Gizis J.~E.,  Burgasser A.~J.,  Faherty J.~K.,  Castro P.~J.,   Shara M.~M.,
  2011a, AJ, 142, 171

\bibitem[\protect\citeauthoryear{Gizis, Troup  \& Burgasser}{Gizis
  et~al.}{2011b}]{Gizis:2011fq}
Gizis J.~E.,  Troup N.~W.,   Burgasser A.~J.,  2011b, ApJL, 736, L34

\bibitem[\protect\citeauthoryear{Gizis, Burgasser, Berger, Williams, Vrba, Cruz
   \& Metchev}{Gizis et~al.}{2013}]{Gizis:2013ik}
Gizis J.~E.,  Burgasser A.~J.,  Berger E.,  Williams P. K.~G.,  Vrba F.~J.,
  Cruz K.~L.,   Metchev S.~A.,  2013, ApJ, 779, 172

\bibitem[\protect\citeauthoryear{Gizis, Burgasser  \& Vrba}{Gizis
  et~al.}{2015}]{Gizis:2015fa}
Gizis J.~E.,  Burgasser A.~J.,   Vrba F.~J.,  2015, AJ, 150, 179

\bibitem[\protect\citeauthoryear{Hawley et~al.,}{Hawley
  et~al.}{2002}]{Hawley:2002jc}
Hawley S.~L.,  et~al., 2002, AJ, 123, 3409

\bibitem[\protect\citeauthoryear{Ho et~al.,}{Ho et~al.}{2017a}]{Ho2017a}
Ho A. Y.~Q.,  et~al., 2017a, \mn@doi [ApJ] {10.3847/1538-4357/836/1/5}, 836, 5

\bibitem[\protect\citeauthoryear{Ho, Rix, Ness, Hogg, Liu  \& Ting}{Ho
  et~al.}{2017b}]{Ho2017b}
Ho A. Y.~Q.,  Rix H.-W.,  Ness M.~K.,  Hogg D.~W.,  Liu C.,   Ting Y.-S.,
  2017b, \mn@doi [ApJ] {10.3847/1538-4357/aa6db3}, 841, 40

\bibitem[\protect\citeauthoryear{Hogg et~al.,}{Hogg et~al.}{2016}]{Hogg2016}
Hogg D.~W.,  et~al., 2016, \mn@doi [ApJ] {10.3847/1538-4357/833/2/262}, 833,
  262

\bibitem[\protect\citeauthoryear{Ivezi{\'{c}}, Connolly, VanderPlas  \&
  Gray}{Ivezi{\'{c}} et~al.}{2014}]{Ivezic2014}
Ivezi{\'{c}} Z.,  Connolly A.~J.,  VanderPlas J.~T.,   Gray A.,  2014,
  {Statistics, data mining, and machine learning in astronomy : a practical
  Python guide for the analysis of survey data}.
Princeton University Press, Princeton

\bibitem[\protect\citeauthoryear{Kellogg, Metchev, Miles-P{\'a}ez  \&
  Tannock}{Kellogg et~al.}{2017}]{Kellogg:2017kh}
Kellogg K.,  Metchev S.~A.,  Miles-P{\'a}ez P.~A.,   Tannock M.~E.,  2017, AJ,
  154, 112

\bibitem[\protect\citeauthoryear{Kendall, Delfosse, Mart{\'\i}n  \&
  Forveille}{Kendall et~al.}{2004}]{Kendall:2004kb}
Kendall T.~R.,  Delfosse X.,  Mart{\'\i}n E.~L.,   Forveille T.,  2004, A{\&}A,
  416, L17

\bibitem[\protect\citeauthoryear{Kendall, Jones, Pinfield, Pokorny, Folkes,
  Weights, Jenkins  \& Mauron}{Kendall et~al.}{2007}]{Kendall:2007fd}
Kendall T.~R.,  Jones H. R.~A.,  Pinfield D.~J.,  Pokorny R.~S.,  Folkes S.,
  Weights D.,  Jenkins J.~S.,   Mauron N.,  2007, MNRAS, 374, 445

\bibitem[\protect\citeauthoryear{Kirkpatrick}{Kirkpatrick}{2005}]{Kirkpatrick:2005cv}
Kirkpatrick J.~D.,  2005, ARA{\&}A, 43, 195

\bibitem[\protect\citeauthoryear{Kirkpatrick et~al.,}{Kirkpatrick
  et~al.}{1999}]{Kirkpatrick:1999ev}
Kirkpatrick J.~D.,  et~al., 1999, ApJ, 519, 802

\bibitem[\protect\citeauthoryear{Kirkpatrick et~al.,}{Kirkpatrick
  et~al.}{2000}]{Kirkpatrick:2000gi}
Kirkpatrick J.~D.,  et~al., 2000, AJ, 120, 447

\bibitem[\protect\citeauthoryear{Kirkpatrick et~al.,}{Kirkpatrick
  et~al.}{2008}]{Kirkpatrick2008}
Kirkpatrick J.~D.,  et~al., 2008, \mn@doi [ApJ] {10.1086/592768}, 689, 1295

\bibitem[\protect\citeauthoryear{Kirkpatrick et~al.,}{Kirkpatrick
  et~al.}{2010}]{Kirkpatrick:2010dc}
Kirkpatrick J.~D.,  et~al., 2010, ApJS, 190, 100

\bibitem[\protect\citeauthoryear{Kirkpatrick et~al.,}{Kirkpatrick
  et~al.}{2011}]{Kirkpatrick:2011ey}
Kirkpatrick J.~D.,  et~al., 2011, ApJS, 197, 19

\bibitem[\protect\citeauthoryear{Kirkpatrick et~al.,}{Kirkpatrick
  et~al.}{2012}]{Kirkpatrick:2012ha}
Kirkpatrick J.~D.,  et~al., 2012, ApJ, 753, 156

\bibitem[\protect\citeauthoryear{Kirkpatrick et~al.,}{Kirkpatrick
  et~al.}{2014}]{Kirkpatrick:2014kv}
Kirkpatrick J.~D.,  et~al., 2014, ApJ, 783, 122

\bibitem[\protect\citeauthoryear{Kirkpatrick et~al.,}{Kirkpatrick
  et~al.}{2019}]{Kirkpatrick:2019kt}
Kirkpatrick J.~D.,  et~al., 2019, ApJS, 240, 19

\bibitem[\protect\citeauthoryear{Kirkpatrick et~al.,}{Kirkpatrick
  et~al.}{2021}]{Kirkpatrick2021}
Kirkpatrick J.~D.,  et~al., 2021, \mn@doi [ApJS] {10.3847/1538-4365/abd107},
  253, 7

\bibitem[\protect\citeauthoryear{Knapp et~al.,}{Knapp
  et~al.}{2004}]{Knapp2004a}
Knapp G.~R.,  et~al., 2004, \mn@doi [AJ] {10.1086/420707}, 127, 3553

\bibitem[\protect\citeauthoryear{Lawrence et~al.,}{Lawrence
  et~al.}{2012}]{Lawrence:2012wh}
Lawrence A.,  et~al., 2012, yCat, II/314, 0

\bibitem[\protect\citeauthoryear{Leggett et~al.,}{Leggett
  et~al.}{2000}]{Leggett:2000ja}
Leggett S.~K.,  et~al., 2000, ApJL, 536, L35

\bibitem[\protect\citeauthoryear{Leggett et~al.,}{Leggett
  et~al.}{2002}]{Leggett:2002cd}
Leggett S.~K.,  et~al., 2002, ApJ, 564, 452

\bibitem[\protect\citeauthoryear{Leggett et~al.,}{Leggett
  et~al.}{2009}]{Leggett:2009jf}
Leggett S.~K.,  et~al., 2009, ApJ, 695, 1517

\bibitem[\protect\citeauthoryear{Leggett et~al.,}{Leggett
  et~al.}{2010}]{Leggett:2010cl}
Leggett S.~K.,  et~al., 2010, ApJ, 710, 1627

\bibitem[\protect\citeauthoryear{Leggett et~al.,}{Leggett
  et~al.}{2012}]{Leggett:2012gg}
Leggett S.~K.,  et~al., 2012, ApJ, 748, 74

\bibitem[\protect\citeauthoryear{Leggett, Morley, Marley, Saumon, Fortney  \&
  Visscher}{Leggett et~al.}{2013}]{Leggett:2013dq}
Leggett S.~K.,  Morley C.~V.,  Marley M.~S.,  Saumon D.,  Fortney J.~J.,
  Visscher C.,  2013, ApJ, 763, 130

\bibitem[\protect\citeauthoryear{Leggett, Morley, Marley  \& Saumon}{Leggett
  et~al.}{2015}]{Leggett:2015dn}
Leggett S.~K.,  Morley C.~V.,  Marley M.~S.,   Saumon D.,  2015, ApJ, 799, 37

\bibitem[\protect\citeauthoryear{Leggett, Tremblin, Saumon, Marley, Morley,
  Amundsen, Baraffe  \& Chabrier}{Leggett et~al.}{2016}]{Leggett:2016fq}
Leggett S.~K.,  Tremblin P.,  Saumon D.,  Marley M.~S.,  Morley C.~V.,
  Amundsen D.~S.,  Baraffe I.,   Chabrier G.,  2016, ApJ, 824, 2

\bibitem[\protect\citeauthoryear{Liebert, Kirkpatrick, Cruz, Reid, Burgasser,
  Tinney  \& Gizis}{Liebert et~al.}{2003}]{Liebert:2003bx}
Liebert J.,  Kirkpatrick J.~D.,  Cruz K.~L.,  Reid I.~N.,  Burgasser A.~J.,
  Tinney C.~G.,   Gizis J.~E.,  2003, AJ, 125, 343

\bibitem[\protect\citeauthoryear{Liu, Fischer, Graham, Lloyd, Marcy  \&
  Butler}{Liu et~al.}{2002}]{Liu:2002fx}
Liu M.~C.,  Fischer D.~A.,  Graham J.~R.,  Lloyd J.~P.,  Marcy G.~W.,   Butler
  R.~P.,  2002, ApJ, 571, 519

\bibitem[\protect\citeauthoryear{Liu et~al.,}{Liu et~al.}{2011}]{Liu:2011hc}
Liu M.~C.,  et~al., 2011, ApJL, 740, L32

\bibitem[\protect\citeauthoryear{Liu et~al.,}{Liu et~al.}{2013}]{Liu:2013gy}
Liu M.~C.,  et~al., 2013, ApJL, 777, L20

\bibitem[\protect\citeauthoryear{Liu, Dupuy  \& Allers}{Liu
  et~al.}{2016}]{LiuDupuy2016}
Liu M.~C.,  Dupuy T.~J.,   Allers K.~N.,  2016, \mn@doi [ApJ]
  {10.3847/1538-4357/833/1/96}, 833, 96

\bibitem[\protect\citeauthoryear{Lodieu, Scholz, McCaughrean, Ibata, Irwin  \&
  Zinnecker}{Lodieu et~al.}{2005}]{Lodieu:2005kd}
Lodieu N.,  Scholz R.-D.,  McCaughrean M.~J.,  Ibata R.,  Irwin M.~J.,
  Zinnecker H.,  2005, A{\&}A, 440, 1061

\bibitem[\protect\citeauthoryear{Lodieu et~al.,}{Lodieu
  et~al.}{2007}]{Lodieu:2007fr}
Lodieu N.,  et~al., 2007, MNRAS, 379, 1423

\bibitem[\protect\citeauthoryear{Lodieu et~al.,}{Lodieu
  et~al.}{2012}]{Lodieu:2012go}
Lodieu N.,  et~al., 2012, A{\&}A, 548, 53

\bibitem[\protect\citeauthoryear{Looper, Kirkpatrick  \& Burgasser}{Looper
  et~al.}{2007}]{Looper:2007ee}
Looper D.~L.,  Kirkpatrick J.~D.,   Burgasser A.~J.,  2007, AJ, 134, 1162

\bibitem[\protect\citeauthoryear{Looper, Gelino, Burgasser  \&
  Kirkpatrick}{Looper et~al.}{2008a}]{Looper2008}
Looper D.~L.,  Gelino C.~R.,  Burgasser A.~J.,   Kirkpatrick J.~D.,  2008a,
  \mn@doi [ApJ] {10.1086/590382}, 685, 1183

\bibitem[\protect\citeauthoryear{Looper et~al.,}{Looper
  et~al.}{2008b}]{Looper:2008hs}
Looper D.~L.,  et~al., 2008b, ApJ, 686, 528

\bibitem[\protect\citeauthoryear{Lucas et~al.,}{Lucas
  et~al.}{2010}]{Lucas:2010iq}
Lucas P.~W.,  et~al., 2010, MNRASL, 408, L56

\bibitem[\protect\citeauthoryear{Lucas et~al.,}{Lucas
  et~al.}{2012}]{Lucas:2012wf}
Lucas P.~W.,  et~al., 2012, yCat, II/316, 0

\bibitem[\protect\citeauthoryear{Luhman et~al.,}{Luhman
  et~al.}{2007}]{Luhman:2007fu}
Luhman K.~L.,  et~al., 2007, ApJ, 654, 570

\bibitem[\protect\citeauthoryear{Luhman et~al.,}{Luhman
  et~al.}{2012}]{Luhman:2012ir}
Luhman K.~L.,  et~al., 2012, ApJ, 760, 152

\bibitem[\protect\citeauthoryear{Mace et~al.,}{Mace et~al.}{2013}]{Mace:2013jh}
Mace G.~N.,  et~al., 2013, ApJS, 205, 6

\bibitem[\protect\citeauthoryear{Manjavacas, Goldman, Reffert  \&
  Henning}{Manjavacas et~al.}{2013}]{Manjavacas:2013cg}
Manjavacas E.,  Goldman B.,  Reffert S.,   Henning T.,  2013, A{\&}A, 560, 52

\bibitem[\protect\citeauthoryear{Marocco et~al.,}{Marocco
  et~al.}{2010}]{Marocco:2010cj}
Marocco F.,  et~al., 2010, A{\&}A, 524, 38

\bibitem[\protect\citeauthoryear{Marocco et~al.,}{Marocco
  et~al.}{2013}]{Marocco:2013kv}
Marocco F.,  et~al., 2013, AJ, 146, 161

\bibitem[\protect\citeauthoryear{Marocco et~al.,}{Marocco
  et~al.}{2015}]{Marocco:2015iz}
Marocco F.,  et~al., 2015, MNRAS, 449, 3651

\bibitem[\protect\citeauthoryear{Mart{\'\i}n et~al.,}{Mart{\'\i}n
  et~al.}{2010}]{Martin:2010cx}
Mart{\'\i}n E.~L.,  et~al., 2010, A{\&}A, 517, 53

\bibitem[\protect\citeauthoryear{Martin et~al.,}{Martin
  et~al.}{2018}]{Martin:2018hc}
Martin E.~C.,  et~al., 2018, ApJ, 867, 109

\bibitem[\protect\citeauthoryear{McMahon, Banerji, Gonzalez, Koposov, Bejar,
  Lodieu, Rebolo  \& {VHS Collaboration}}{McMahon
  et~al.}{2013}]{McMahon:2013vw}
McMahon R.~G.,  Banerji M.,  Gonzalez E.,  Koposov S.~E.,  Bejar V.~J.,  Lodieu
  N.,  Rebolo R.,   {VHS Collaboration} 2013, The Messenger, 154, 35

\bibitem[\protect\citeauthoryear{Mugrauer, Seifahrt, Neuh{\"a}user  \&
  Mazeh}{Mugrauer et~al.}{2006}]{Mugrauer:2006iy}
Mugrauer M.,  Seifahrt A.,  Neuh{\"a}user R.,   Mazeh T.,  2006, MNRASL, 373,
  L31

\bibitem[\protect\citeauthoryear{Ness, Hogg, Rix, Ho  \& Zasowski}{Ness
  et~al.}{2015}]{Ness2015}
Ness M.,  Hogg D.~W.,  Rix H.~W.,  Ho A.~Y.,   Zasowski G.,  2015, \mn@doi
  [ApJ] {10.1088/0004-637X/808/1/16}, 808, 16

\bibitem[\protect\citeauthoryear{Ness, Hogg, Rix, Martig, Pinsonneault  \&
  Ho}{Ness et~al.}{2016}]{Ness2016}
Ness M.,  Hogg D.~W.,  Rix H.-W.,  Martig M.,  Pinsonneault M.~H.,   Ho A.
  Y.~Q.,  2016, \mn@doi [ApJ] {10.3847/0004-637x/823/2/114}, 823, 114

\bibitem[\protect\citeauthoryear{Pe{\~n}a~Ram{\'\i}rez, Zapatero~Osorio  \&
  B{\'e}jar}{Pe{\~n}a~Ram{\'\i}rez et~al.}{2015}]{PenaRamirez:2015id}
Pe{\~n}a~Ram{\'\i}rez K.,  Zapatero~Osorio M.~R.,   B{\'e}jar V. J.~S.,  2015,
  A{\&}A, 574, A118

\bibitem[\protect\citeauthoryear{Phan-Bao et~al.,}{Phan-Bao
  et~al.}{2008}]{PhanBao:2008kz}
Phan-Bao N.,  et~al., 2008, MNRAS, 383, 831

\bibitem[\protect\citeauthoryear{Pineda, Hallinan, Kirkpatrick, Cotter, Kao  \&
  Mooley}{Pineda et~al.}{2016}]{Pineda:2016ku}
Pineda J.~S.,  Hallinan G.,  Kirkpatrick J.~D.,  Cotter G.,  Kao M.~M.,
  Mooley K.,  2016, ApJ, 826, 73

\bibitem[\protect\citeauthoryear{Pinfield et~al.,}{Pinfield
  et~al.}{2008}]{Pinfield:2008jx}
Pinfield D.~J.,  et~al., 2008, MNRAS, 390, 304

\bibitem[\protect\citeauthoryear{Pinfield et~al.,}{Pinfield
  et~al.}{2012}]{Pinfield:2012hm}
Pinfield D.~J.,  et~al., 2012, MNRAS, 422, 1922

\bibitem[\protect\citeauthoryear{Radigan, Lafreniere, Jayawardhana  \&
  Doyon}{Radigan et~al.}{2008}]{Radigan:2008jd}
Radigan J.,  Lafreniere D.,  Jayawardhana R.,   Doyon R.,  2008, ApJ, 689, 471

\bibitem[\protect\citeauthoryear{Rayner, Toomey, Onaka, Denault, Stahlberger,
  Vacca, Cushing  \& Wang}{Rayner et~al.}{2003}]{Rayner2003}
Rayner J.,  Toomey D.,  Onaka P.,  Denault A.,  Stahlberger W.,  Vacca W.,
  Cushing M.,   Wang S.,  2003, \mn@doi [PASP] {10.1086/367745}, 115, 362

\bibitem[\protect\citeauthoryear{Reid, Kirkpatrick, Gizis, Dahn, Monet,
  Williams, Liebert  \& Burgasser}{Reid et~al.}{2000}]{Reid:2000iw}
Reid I.~N.,  Kirkpatrick J.~D.,  Gizis J.~E.,  Dahn C.~C.,  Monet D.~G.,
  Williams R.~J.,  Liebert J.,   Burgasser A.~J.,  2000, AJ, 119, 369

\bibitem[\protect\citeauthoryear{Reid, Cruz, Kirkpatrick, Allen, Mungall,
  Liebert, Lowrance  \& Sweet}{Reid et~al.}{2008}]{Reid:2008fz}
Reid I.~N.,  Cruz K.~L.,  Kirkpatrick J.~D.,  Allen P.~R.,  Mungall F.,
  Liebert J.,  Lowrance P.,   Sweet A.,  2008, AJ, 136, 1290

\bibitem[\protect\citeauthoryear{Sahlmann, Lazorenko, S{\'e}gransan,
  Mart{\'\i}n, Mayor, Queloz  \& Udry}{Sahlmann et~al.}{2014}]{Sahlmann:2014hu}
Sahlmann J.,  Lazorenko P.~F.,  S{\'e}gransan D.,  Mart{\'\i}n E.~L.,  Mayor
  M.,  Queloz D.,   Udry S.,  2014, A{\&}A, 565, A20

\bibitem[\protect\citeauthoryear{Schneider, Cushing, Kirkpatrick, Mace, Gelino,
  Faherty, Fajardo-Acosta  \& Sheppard}{Schneider
  et~al.}{2014}]{Schneider:2014jd}
Schneider A.~C.,  Cushing M.~C.,  Kirkpatrick J.~D.,  Mace G.~N.,  Gelino
  C.~R.,  Faherty J.~K.,  Fajardo-Acosta S.,   Sheppard S.~S.,  2014, AJ, 147,
  34

\bibitem[\protect\citeauthoryear{Schneider et~al.,}{Schneider
  et~al.}{2015}]{Schneider:2015bx}
Schneider A.~C.,  et~al., 2015, ApJ, 804, 92

\bibitem[\protect\citeauthoryear{Scholz}{Scholz}{2010}]{Scholz:2010cy}
Scholz R.-D.,  2010, A{\&}A, 515, 92

\bibitem[\protect\citeauthoryear{Scholz \& Meusinger}{Scholz \&
  Meusinger}{2002}]{Scholz:2002by}
Scholz R.-D.,  Meusinger H.,  2002, MNRAS, 336, L49

\bibitem[\protect\citeauthoryear{Scholz, Bihain, Schnurr  \& Storm}{Scholz
  et~al.}{2011}]{Scholz:2011gs}
Scholz R.-D.,  Bihain G.,  Schnurr O.,   Storm J.,  2011, A{\&}A, 532, L5

\bibitem[\protect\citeauthoryear{Simons \& Tokunaga}{Simons \&
  Tokunaga}{2002}]{Simons:2002hh}
Simons D.~A.,  Tokunaga A.~T.,  2002, PASP, 114, 169

\bibitem[\protect\citeauthoryear{Skrutskie et~al.,}{Skrutskie
  et~al.}{2006}]{Skrutskie2006}
Skrutskie M.~F.,  et~al., 2006, \mn@doi [AJ] {10.1086/498708}, 131, 1163

\bibitem[\protect\citeauthoryear{Smart et~al.,}{Smart
  et~al.}{2018}]{Smart:2018en}
Smart R.~L.,  et~al., 2018, MNRAS, 481, 3548

\bibitem[\protect\citeauthoryear{Spiegel, Burrows  \& Milsom}{Spiegel
  et~al.}{2011}]{Spiegel2011}
Spiegel D.~S.,  Burrows A.,   Milsom J.~A.,  2011, \mn@doi [ApJ]
  {10.1088/0004-637X/727/1/57}, 727, 1

\bibitem[\protect\citeauthoryear{Strauss et~al.,}{Strauss
  et~al.}{1999}]{Strauss:1999iw}
Strauss M.~A.,  et~al., 1999, ApJL, 522, L61

\bibitem[\protect\citeauthoryear{Thompson et~al.,}{Thompson
  et~al.}{2013}]{Thompson:2013kv}
Thompson M.~A.,  et~al., 2013, PASP, 125, 809

\bibitem[\protect\citeauthoryear{Tinney, Burgasser  \& Kirkpatrick}{Tinney
  et~al.}{2003}]{Tinney:2003eg}
Tinney C.~G.,  Burgasser A.~J.,   Kirkpatrick J.~D.,  2003, AJ, 126, 975

\bibitem[\protect\citeauthoryear{Tinney, Burgasser, Kirkpatrick  \&
  McElwain}{Tinney et~al.}{2005}]{Tinney:2005hz}
Tinney C.~G.,  Burgasser A.~J.,  Kirkpatrick J.~D.,   McElwain M.~W.,  2005,
  AJ, 130, 2326

\bibitem[\protect\citeauthoryear{Tokunaga, Simons  \& Vacca}{Tokunaga
  et~al.}{2002}]{Tokunaga:2002ex}
Tokunaga A.~T.,  Simons D.~A.,   Vacca W.~D.,  2002, PASP, 114, 180

\bibitem[\protect\citeauthoryear{Tsvetanov et~al.,}{Tsvetanov
  et~al.}{2000}]{Tsvetanov:2000cg}
Tsvetanov Z.~I.,  et~al., 2000, ApJL, 531, L61

\bibitem[\protect\citeauthoryear{Vrba et~al.,}{Vrba et~al.}{2004}]{Vrba:2004ee}
Vrba F.~J.,  et~al., 2004, AJ, 127, 2948

\bibitem[\protect\citeauthoryear{Warren et~al.,}{Warren
  et~al.}{2007}]{Warren:2007kw}
Warren S.~J.,  et~al., 2007, MNRAS, 381, 1400

\bibitem[\protect\citeauthoryear{Wilson, Miller, Gizis, Skrutskie, Houck,
  Kirkpatrick, Burgasser  \& Monet}{Wilson et~al.}{2003}]{Wilson:2003tk}
Wilson J.~C.,  Miller N.~A.,  Gizis J.~E.,  Skrutskie M.~F.,  Houck J.~R.,
  Kirkpatrick J.~D.,  Burgasser A.~J.,   Monet D.~G.,  2003, in Mart{\'\i}n
  E.~L.,  ed., IAU Symp. 211, Brown Dwarfs. ASP, San Francisco, CA, p.~197

\bibitem[\protect\citeauthoryear{Wright et~al.,}{Wright
  et~al.}{2010}]{Wright2010}
Wright E.~L.,  et~al., 2010, \mn@doi [AJ] {10.1088/0004-6256/140/6/1868}, 140,
  1868

\bibitem[\protect\citeauthoryear{Wright et~al.,}{Wright
  et~al.}{2013}]{Wright:2013bo}
Wright E.~L.,  et~al., 2013, AJ, 145, 84

\bibitem[\protect\citeauthoryear{van Leeuwen}{van
  Leeuwen}{2007}]{vanLeeuwen:2007dc}
van Leeuwen F.,  2007, A{\&}A, 474, 653

\makeatother
\end{thebibliography}

\bsp	
\label{lastpage}
\end{document}